\documentclass[12pt,a4paper]{article}
\usepackage[utf8]{inputenc}
\usepackage{amsmath}
\usepackage[table]{xcolor}
\usepackage{xcolor}
\usepackage{multirow}
\definecolor{myOrange}{RGB}{44,14,14}
\usepackage{amsfonts}
\usepackage{afterpage}
\usepackage{amssymb}
\usepackage{array}
\usepackage{tabularx}
\newcolumntype{P}[1]{>{\centering\arraybackslash}p{#1}}
\newcolumntype{M}[1]{>{\centering\arraybackslash}m{#1}}

\usepackage{dsfont}
\usepackage{hyperref}
\usepackage{relsize}
\usepackage{graphicx}
\usepackage{grffile}
\usepackage[section]{placeins}
\usepackage{sectsty}
\usepackage{cancel}
\sectionfont{\fontsize{15}{15}\selectfont}
\subsectionfont{\fontsize{13}{15}\selectfont}
\usepackage{marginnote}
\usepackage[obeyFinal,textsize=tiny,color=yellow]{todonotes}
\usepackage{authblk} 

\usepackage{siunitx}

\usepackage{caption}
\newcolumntype{P}[1]{>{\centering\arraybackslash}p{#1}}
\usepackage{enumerate}
\usepackage[style=authoryear-icomp,maxbibnames=9,maxcitenames=2,uniquelist=false, backend=biber, sortcites, natbib]{biblatex}

\defbibenvironment{bibliography}
  {\begin{enumerate}}
  {\end{enumerate}}
  {\item}

\bibliography{library-Adaptation, library-Adaptation}
\usepackage[english]{babel}
\bibdata{library}

\newcommand{\note}[1]{\marginpar{\begin{flushleft}
\end{flushleft}}}

\title{A FRAMEWORK TO QUANTIFY ADAPTATION TO MULTIPLE DRIVERS}
\author[]{Emily Quiroga\thanks{Corresponding author: emily.quiroga-gomez@uni-hamburg.de}}
\author[]{Benjamin Blanz\thanks{benjamin.blanz@uni-hamburg.de}}
\affil[]{Research Unit Sustainability and Climate Risk, University of Hamburg, Grindelberg 5, 20144 Hamburg, Germany}
\date{}

\allowdisplaybreaks

\begin{document}

\maketitle

\begin{abstract}
We develop an analytical framework to assess the adaptations  in a coupled ecological-economic system and apply it to a bio-economic model. Our framework allows us to quantify the impact of multiple drivers on a coupled ecological-economic system, while distinguishing between adaptation and sensitivities to positive and negative exposures. This distinction allows us to differentiate between drivers that improve and decrease well-being. Our findings provide insight into how to focus resources to counteract negative or enhance positive impacts. We apply this framework to a bio-economic model calibrated to the North Sea flatfish fishery. We quantify the adaptations, sensitivities and total impact of fishers' profits to multiple drivers and identify among which of them fishers adapt the most. This work forms a bridge between the multidisciplinary area of adaptability and the bio-economic modelling domain,increasing the understanding and knowledge regarding the measure of adaptation.
\end{abstract}

\setlength{\parskip}{5mm}






\newpage
\section{Introduction}
The interaction between humans and nature involves multiple complexities and feedback affected by numerous sets of socioeconomic and ecological drivers. The impact of these drives on the coupled system happens in expected and unexpected ways. Here we focus on the human behaviour and how humans adapt to these challenges from the environment or society. The analysis of adaptation is rooted in a multidisciplinary field with various approaches guided by their respective scientific background. These are the risk hazard approach, human/political ecology approach, and the ecological resilience approach \citep{Berrouet2018}. The political ecology view measures adaptation based on resources and social variables such as capital, education, income, and, social capital. The ecological resilience view argues that adaptation is not only about resources but about actions that sustain pathways of a socio-ecological system \citep{Folke2016}. In this study, we offer an alternative view of adaptation using economic theory as a background. This alternative offers us a way to derive the optimal adaptive response for each driver and to distinguish the adaptation response among positive and negative impacts allowing comparison among multiple drivers. 

Adaptation can be separated into three stages, adaptation before, during, and after an impact. The first is called absorptive adaptation usually reducing the risk and exposure to drivers, while the latter is related to long-term responses, where adjustments become habitual \citep{Hufschmidt2011}. The adaptive capacity is considered a potential adaptation, i.e., before an impact. In the human/political ecology approach the adaptive capacity is measured using indicators such as access to assets, livelihoods, or governance and institutional aspects \citep{Reed2013,Whitney2017, Serfilippi2018}. These indicators, although useful, describe the access to resources and not if those resources would be used when changes in drivers are experienced. Moreover, they describe general adaptations to deal with any harm and conceal driver-specific adaptations \citep{Thiault2019a,McDowell2012}. Here we present a framework that estimates the impacts of multiple drivers on a coupled ecological-economic system allowing a comparison among them.

A contribution of our framework is to describe the optimal adaptive response for each driver in a coupled ecological-economic system. We operationalize the concept by \citet{Ionescu2009} where an adaptive capacity is an action in which the performance of the system is preferable to the performance without it. Sometimes the general adaptation measures before an impact can or can not be effective during the impact. I.e, communities, entities, or individuals can have access to resources (subsidies, insurance, or education), but due to institutional or governmental reasons, these can not be used in the midst of an impact. Our framework focuses on adaptation instead of adaptive capacity. The first refers to an actual or expected behaviour, and the second to the potential of adaptation. Our framework elicits the system's performance regarding a change in adaptation before, during, and after a change in drivers. This distinction allows us to disentangle the optimal adaptive response during an impact, and the performance of the system after it. We are interested in the degree to which the optimal adaptive response could mitigate an impact, rather than the time the response takes. These responses, however, can take a shorter or longer time depending on the phenomena of the system analyzed.


As a proof of concept, the framework is applied to a calibrated bio-economic multi-species model of the North Sea flatfish fishery. We determine the optimal adaptive effort responses of fishers' profits to changes in returns to effort, stock harvesting efficiency and wages. We show the impacts of adaptation through effort on the quantities harvested and utility of the system. Our work add to the already existing models of this mixed flatfish fishery by focusing on the adaptation through effort to changes in multiple drivers \citep{Nielsen2018, Prellezo2012}. The stylish nature of the model we use allow us to understand the mechanisms underlying the adaptation and how these affect the quantities produced in the market and the utility of the system. Our results indicate that fish quantities are mostly affected by adaptation due to changes in returns to effort followed by stock harvesting efficiency and wages. The utility is mostly affected by changes in wages, however the role of effort adaption in influencing utility is very low. By considering multiple drivers our framework allows to identify trade-offs among impacts on a given property of the system.

\section{Adaptation across analytic approaches}


Definitions of adaptation have been analysed among a diversity of disciplines. Analytic approaches such as ecological resilience, human/political ecology, and risk-hazard have different definitions of adaptation and adaptive capacity \footnote{This classification of schools of thought is based on \citet{Berrouet2018}, however,  \citet{Adger2006}  show other distinctions such as  `the vulnerability as an absence of entitlements', `Natural Hazards', and `Human/Political Ecology'. \citet{Hufschmidt2011} also classifies the 'human ecologist school', `structural view', and `resilience school'. }. The ecological resilience approach conceives resilience as a system property.  It is the system's capacity to self-organize and adapt in the face of ongoing change in a way that sustains the system in certain stability \citep{Folke2016}. In the `Human/Political Ecology' approach a difference between `adjustment' and `adaptation' is made. Adjustments are purposeful actions, such as building a dam or structure to resist earthquakes. Adaptation is regarded as a process of co-evolution between an organism and its environment in a long-term response \citep{Hufschmidt2011}. 	In the approach combining hazards and human/political ecology definitions such as the given by the  Intergovernmental Panel of Climate Change (IPCC) define adaptation as ``the process of adjustment to actual or expected climate and its effects to moderate harm or exploit beneficial opportunities" \citep{p.43, IPCC2022}. Also,  \citet{Whitney2017} ``refers to the latent ability of a system to respond proactively and positively to stressors or opportunities" \citep{Whitney2017}.    

The definitions of adaptation above all cover both actions to moderate harms and to exploit benefits, however, their measures do not always show this distinction. Measures following the human/ecology approach are mostly directed to reduce harm, they use socio-ecological indicators, institutional analysis, social experiments, and community-based approaches as a way to measure adaptations \citep{Whitney2017}. The ecological resilience approach presents measures of adaptation, that usually contemplate the existence of thresholds. For instance, \citet{Luers2003} quantifies adaptive capacity as the difference in vulnerability under existing conditions and modified conditions. Here, a system is described as a function of well-being ($W$), threshold ($W_0$), and a stressor ($X$). Vulnerability is then measured as the sensitivity regarding a threshold ($V=f(\frac{\lvert \partial W/ \partial X \lvert}{W/W_0})$). Furthermore, \citet{Grafton2019} shows a measure of resilience with three main characteristics resistance, recovery, and robustness. Resistance is the system's ability to actively change while maintaining its system performance following one or more adverse events. Recovery is the time a system's performance needs to recover a desired functionality after an adverse event, and robustness is the system's probability to maintain its identity and not cross an undesirable threshold after an adverse event. These measures, however, do not cover adaptation with a positive impact and do not distinguish adaptation with both positive and negative impacts.

The measures described by \citet{Luers2003} and \citet{Grafton2019} require a definition of a threshold in a system, however, in many systems, this threshold can not be defined or simply does not exist. Our framework adds to the literature by quantifying not only negative impacts (that may drive the system close to a threshold) but also positive impacts which enhance the system's performance.  We focus on quantifying the optimal adaptive response. We operationalize the concept of adaptation defined by \citet{Ionescu2009} and show how this optimal adaptive response may change with a positive and negative impact. \citet{Ionescu2009} define a framework in which a system is described as a function of the state of the system ($x$), a given input ($e$), and an adaptive action ($u$). They define an optimal action ($u \in U$) such that $f(x,e,u)$ is optimal. However, sometimes there is no complete knowledge of $f$ and they define adaptation as an action where the performance of the system within that action is preferred to the performance of the system without it. This is important because the optimal action serves as a point of reference for the best scenario to be achieved during an impact so that efforts and resources can be well directed.

In general, most of the definitions focus on adaptation as an ongoing process. The states of this process are defined differently according to the approach\footnote{\citet{Hufschmidt2011} mentions the term `adaptation' as the process of learning, anticipating, modifying, preparing, and planning. She distinguishes adaptive activities for households in a stage of mitigation, preparation, or recovery. \citet{Bene2012} departing from the resilience approach distinguishes absorptive, adaptive, and transformative capacities. The absorptive capacity reduces the risk of exposure to shocks absorbing the impact in the short term. While adaptive and transformative capacities are long-term responses to socio-economic and environmental challenges\citep{Serfilippi2018}.}, but in general, they refer to actions before, during, and after an impact. Here we focus on adaptation during an impact, also called `reacting action', `response', or `coping capacity' as an action during crisis \citep{Hufschmidt2011}. In practice, we only observe the state of the system before and after an impact, the latter already embeds the adaptive response, i.e., the effective resources or abilities used to cope with the impact.  Our framework aims to quantify and disentangle this impact. We add to the literature by identifying the magnitude that the adaptive response can mitigate harmful impacts or can enhance beneficial impacts on the system. We describe four types of adaptive response the first two evaluate the absolute and marginal changes in the system's well-being function driven by a change in endogenous adaptation response. The third measures the rate at which the adaptive response changes due to marginal changes in the driver. The fourth measures how the adaptive response changes itself given marginal changes in the driver.

\section{Framework to Quantify Impacts and Adaptation in an coupled ecological-economic system}

Our approach to assess adaptation is based on the approach developed by \citet{Ionescu2009}, who developed a formal framework of vulnerability to climate change. Vulnerability is defined using mathematical concepts independent of any knowledge domain and applicable to any system under consideration. Their vulnerability definition is based on the Intergovernmental Panel on Climate Change (IPCC). This definition states that the vulnerability to climate change is a function of the character, magnitude and rate of climate variation to which a system is exposed, its sensitivity and its adaptive capacity" \citep{Ionescu2009}. The vulnerability depends on the differences in exposure to the various direct effects of climate change which lead to different sensitivities and hence generating differential potential impacts on the system. The adaptive capacity is defined as  the ability of a system to adjust to climate change to moderate potential damages, to take advantage of opportunities or to cope with the consequences. 

We consider the term adaptation in a broad sense, i.e, actions within the system taken not only to mitigate harmful impacts but also to enhance the positive impacts. A coupled ecological-economic system is exposed to multiple drivers that generate an increase/decrease of the performance in the system and adaptation aims to mitigate/enhance those impacts. In the human/political approach the differentiation between 'adjustment' and 'adaptation' lies in the temporal distinction, where the first are purposeful actions to adapt and the latter refers to long-term response, where adjustments become a part of society's habitus \citep{Hufschmidt2011} \footnote{There is a temporal distinction between adjustments and adaptation which is difficult to define since the point where adjustments evolve into society's habitus is hazy  \citep{Hufschmidt2011}}. In our framework the term adaptation refers to the long term response and addresses the issue of identifying the degree to which this response mitigates/enhances the impact of a driver. 

Following the human/political ecology approach and using economic theory our framework aims to quantify the best case potential adaptation response of a system to a specific driver. This measurement can help decision-makers to have a reference point of the magnitude of the adaptive response that could be achieved by performing certain activities to mitigate/enhance the impact of multiple drivers.  Our framework is designed to answer the question \textit{the adaptation of what to what?}. \citet{Ionescu2009} state that vulnerability and adaptive capacity are relative properties, it is the adaptation \textit{of} something \textit{to} something. Hence, our methodology encompasses two steps. Identification of (i) the system property under analysis (\textit{of what}), and (ii) the driver (\textit{to what}). The system property refers to the specific aspect of the coupled ecological-economic system considered. For example, in our case study, we investigate the \textit{adaptation of fisher profits to changes in e.g. wages} and other drivers. In the following, we present the formal definitions of drivers, exposure, sensitivity, adaptation, and total impact (TEI).

\subsection{Formalisation}

\subsection*{Drivers}
We define $\theta = ( \theta_1, \dots,\theta_D )$ as the vector of $D$ drivers of the coupled ecological-economic system, for which the researcher wishes to investigate the impacts on a specific system property. For instance, $\theta_d$ can represent the value of an input in a certain process affecting the system property. All drivers are considered to be exogenous depending on the boundaries of the system investigated.

\subsection*{System Property}
We define $\psi(\theta)$ as the property of the system under investigation. This property can be related to the economic, ecological or social side of the system depending on the research question given by the researcher. Multiple properties can be also evaluated separately, case in which $\psi(\theta)$ becomes a vector valued function with $P$ properties . 

\subsection*{Adaptation}
In addition to the drivers, the system property ($\psi$) also depends on $\tau(\theta)$, which corresponds to the endogenous behaviours in response to the drivers $\theta$. We define $\tau(\theta)=(\tau_1(\theta),\dots,\tau_M(\theta))$ as the $M$ adaptation variables of actors within the system. A system can have a single or multiple adaptation variables. The optimal $\tau^{*}(\theta)$ is the value that maximizes the system property $\psi(\theta)$.

\begin{align}
\tau^{*}(\theta) = \operatorname*{argmax}_\tau \psi(\tau, \theta)
\label{eq:optAdapt}
\end{align}

\subsection{Exposure}
Exposure to changes in drivers, or simply exposure, is the magnitude of change in any drivers affecting the system property. For determining adaptions, the source of these events is not relevant, only their magnitude. This can either be evaluated for the entire vector of drivers or individual drivers. Usually, exposure is dependent on impacting a particular part of a system. In our definition a system property can be exposed but not affected, case in which the sensitivity (how affected the system property is by changes in drivers) would be zero. For instance, if the system property of a coupled ecological-economic system is a measure of a community's well-being there could be changes in drivers which do not affect the community's well-being.

\begin{align}
E(\theta,{^0\theta})&=\theta- {^0\theta} \\
E_d(\theta_d,{^0\theta_d} ) &= {\theta_d} - {^0\theta_d} 
\label{eq:exposure1}
\end{align}

Each $E_d( \theta_d,{^0\theta_d})$ depends on the magnitude of change in the driver $d$, where ${^0\theta_d}$  is the original value of the driver, and ${\theta_d}$ is the new state (Eq. \eqref{eq:exposure1}). The vector ${^0\theta}$ contains the initial values of all drivers. ${\theta_d}$ can be higher or lower than the initial state, resulting in a positive or negative exposure. If changes in a single driver, e.g. $\theta_d$, are evaluated the vector of exposure contains zeros in all positions except for the change in $\theta_d$ in the $d$th position 
($E(\theta,{^0\theta}) = (0,\dots,\theta_d - {^0\theta_d},\dots,0)$).

\subsection{Sensitivity}

The sensitivity is the degree to which the system property is affected either adversely or beneficially by exposure to changes in drivers \citep{IPCC2001}, given their initial values and excluding any adaptation. The sensitivity to a given level of exposure may vary depending on the system property under analysis. We interpret this as the change on the system property given by a change in the driver (Eq.\eqref{eq:sensitivity1}). 

We define continuous and absolute sensitivities regarding the impact on the system property. The absolute measure is useful when investigating the total impact considering the range of exposure levels of the driver. Marginal sensitivities show the rate of change in the system property given by a marginal change in driver.

\subsubsection*{Absolute}
In Eq.\eqref{eq:sensitivity1} we evaluate the system property ($\psi$) in two points, at the initial state of the drivers $\psi ({^0\theta}, \tau(^0\theta))$ and at the new state $\psi({\theta}, \tau(^0\theta))$, with no change in adaptation $\tau(^0\theta)$. Depending on the data availability Eq. \eqref{eq:sensitivity1} can be evaluated in many values for each  driver considered. For each property the sensitivity $\psi_p$ is measured by the difference in the system property induced by the exposure, without adaptation. The absolute sensitivity can have positive or negative values, it depends on the effect of the driver on the system property. I.e., if $\psi_p(\theta)$  is greater than the value of the system property at the initial state ($\psi_p(^0\theta)$) then the sensitivity with respect to that property $S_p(\theta,{^0\theta})$ is positive, otherwise it is negative. If the change of the driver affects the system property adversely $\theta_d$ is considered a stressor, otherwise a benefactor.

\begin{align}
S({\theta},{^0\theta}) &= \psi({\theta,\tau(^0\theta})) - \psi({^0\theta,\tau(^0\theta)}) \nonumber \\
 &=(S_1({\theta},{^0\theta}),\dots, S_P({\theta},{^0\theta}) ) \nonumber \\
 &= (\psi_1({\theta}, \tau({^0\theta})) - \psi_1({^0\theta}, \tau({^0\theta})),\dots, \psi_P({\theta}, \tau({^0\theta})) - \psi_P({^0\theta}, \tau({^0\theta})) )
\label{eq:sensitivity1}
\end{align}

\subsubsection*{Marginal}
The marginal sensitivities evaluate impacts on the system property from marginal changes in a driver at a given point. It measures the impact of a marginal increase in exposure from this point disregarding non-linearities in responses to larger exposure levels. This is relevant when making policy choices that are robust to random shocks \footnote{We follow \citet{Gallopin2006} who defines sensitivity as change in the transformation of the system with respect to a change in the perturbation.}. In the case that multiple properties and drivers are evaluated simultaneously, the marginal form is the jacobian of Eq. \eqref{eq:sensitivity1}. The entries $s_{pd}$ give the marginal sensitivity of property $p$ to a change in driver $d$. In the case when a single property is considered $P=1$ the Jacobian matrix collapses to a vector of partial derivatives.

\begin{equation}
s({\theta},{^0\theta}) = \left(\begin{array}{ccc}
\dfrac{\partial S_{1}({\theta},{^0\theta})}{\partial \, \theta_{1}} & \cdots & \dfrac{\partial S_{1}({\theta},{^0\theta})}{\partial \, \theta_{D}} \\
\vdots & \ddots & \vdots \\
\dfrac{\partial S_{P}({\theta},{^0\theta})}{\partial \, \theta_{1}} & \cdots & \dfrac{\partial S_{P}({\theta},{^0\theta})}{\partial \, \theta_{D}}
\end{array}\right)
\end{equation}

\begin{equation}
s_{pd}(\theta,{^0\theta}) = \frac{\partial S_p({\theta},{^0\theta})}{\partial \, \theta_d} = \frac{\partial  \psi_p(\theta,\tau(^0\theta))}{\partial \, \theta_d} 
\label{eq:sensitivity2}
\end{equation}

\subsection{Adaptation}

We define adaptation as the ability of an element within a coupled ecological-economic system to adjust to changing external drivers. Adaptation moderates harm or exploits beneficial opportunities \citep{IPCC2014a}.
The system properties ($\psi$) also depend on $\tau(\theta)$, which corresponds to the endogenous behaviours in response to the drivers $\theta$. The adaption measures how much an optimal response to a change in the drivers can improve the system property, compared to the outcome without an adaptation (Eq. \eqref{eq:adaptiveDis}). Additionally, we also measure the amount of change in the endogenous behaviour that is necessary to achieve the optimal adaptation.

\subsubsection*{Absolute}

Eq.\eqref{eq:adaptiveDis} shows the difference between the system property evaluated with an endogenous response to the drivers $\tau({\theta})$, and the initial behaviour $\tau({^0\theta})$ with no response. In the case of multiple behaviour variables $\tau(\theta)$ is a vector. For instance, to assess the adaptive response \textit{of} a community's well being \textit{to} climate change, $\psi_p(\theta, \tau(\theta))$ corresponds to the system property under evaluation, i.e.,  community's well-being, a measure of the outcome. $\theta$ are drivers affected by climate change, and $\tau(\theta)$ reflects the community's actions affecting their well-being. $\tau(\theta)$ changes in response to the drivers $\theta$. The community's well-being $\psi(\theta)$ can be some measure of utility, socio-economic or financial characteristics. The adaptive capacity is the  benefit to the community of adapting to climate change, determined as the difference in well-being in the community before and after adaptation. $\psi$ is evaluated at the new value of the driver $\theta$, and there is only change in $\tau$. If $\psi(\theta, \tau(\theta))$ is a vector of multiple properties being evaluated ${^aA}({\theta},{^0\theta})$ is a vector valued function, where each entry corresponds to the changes in one of the properties.

\begin{align}
{^aA}({\theta},{^0\theta}) &= \psi({\theta}, \tau({\theta})) -\psi({\theta}, \tau({^0\theta})) \nonumber \\
 &=({^aA}_1({\theta},{^0\theta}),...., {^aA}_P({\theta},{^0\theta}) ) \nonumber \\
 &= (\psi_1({\theta}, \tau({\theta})) - \psi_1({\theta}, \tau({^0\theta})),...., \psi_P({\theta}, \tau({\theta})) - \psi_P({\theta}, \tau({^0\theta})) )
\label{eq:adaptiveDis}
\end{align}

The change in behaviour in order to adapt is the difference in $\tau(\theta)$ due to the change in $\theta$ (Eq.\eqref{eq:adaptiveEffortFramework}).

\begin{align}
{^cA}({\theta},{^0\theta}) &= \tau({\theta}) - \tau({^0\theta}) \nonumber \\
 &= (\tau_1({\theta}) - \tau_1({^0\theta}),....,
\tau_M({\theta}) - \tau_M({^0\theta})) 
\label{eq:adaptiveEffortFramework}
\end{align}

\subsubsection*{Marginal}

We consider three marginal measures for adaptive capacity. First, the marginal version Eq. \ref{eq:adaptiveDis} is the Jacobian with the elements ${^aa_{pd}}(\theta)$. The entry ${^aa_{pd}}(\theta)$ represents the change in the mitigation of sensitivity of the system property $p$ given by a change in the adaptation behaviour ($\tau$) due to a marginal change of the driver $d$ (Eq. \ref{eq:adaptiveDis1}). Second, as the marginal adaptive capacity of Eq. (\ref{eq:adaptiveDis1}) is zero in the zero exposure case we also consider the second derivatives of Eq. (\ref{eq:adaptiveDis}). The elements ${^ba_{pd}}(\theta)$ present the second partial derivatives of Eq. \ref{eq:adaptiveDis}. This is the curvature of the adaptive capacity, the rate at which ${^aa_{pd}}(\theta)$ changes due to a marginal change in $\theta_d$. Third, ${^ca_{md}}(\theta)$ is the marginal measure of $^cA(\theta, ^0\theta)$. It shows the marginal optimal change of adaptation behaviour in $\tau_{m}$, given a marginal increase in driver $d$(Eq. \ref{eq:adaptiveCon}). Notice that ${^ca_{md}}(\theta)$ measures changes in the ability of adaptation while ${^aa_{pd}}(\theta)$ and ${^ba_{pd}}(\theta)$ are about changes in the benefit of adaptation.

\begin{align}
{^aa_{pd}}(\theta)& = \frac{\partial A_p({\theta},{^0\theta})}{\partial \, \theta_d} \nonumber \\
\label{eq:adaptiveDis1} \\
& = \frac{\partial \psi_p({\theta}, \tau({\theta}))}{\partial \, \theta_d} - \frac{\partial \psi_p({\theta}, \tau({^0\theta}))}{\partial \, \theta_d} \nonumber \\
& = v_{pd}(\theta) - s_{pd}(\theta)
\label{eq:adaptiveDis1a} \\
&\phantom{=}\nonumber \\
{^ba_{pd}}(\theta) & = \frac{\partial^2 {A_{pd}({\theta}},{^0\theta})}{\partial^2 \, {\theta_d^2}} \nonumber \\
& = \frac{\partial^2 \psi_{pd}({\theta}, \tau({\theta}))}{\partial^2 \, \theta^2_d} - \frac{\partial^2 \psi_{pd}({\theta}, \tau({^0\theta}))}{\partial^2 \, \theta_d^2} \nonumber \\
& = \frac{\partial v_{pd}(\theta)}{\partial \theta_d} - \frac{\partial s_{pd}(\theta)}{\partial \theta_d} 
\label{eq:adaptiveDis2} \\
&\phantom{=}\nonumber \\
{^ca_{md}}(\theta) &= \frac{\partial \tau_m(\theta)}{\partial \, \theta_d} 
\label{eq:adaptiveCon} 
\end{align}

For instance, to assess the adaptive capacity \textit{of} a community's well being \textit{to} climate change, $\psi(\theta, \tau(\theta))$ represents a single measure of community's well being affected by climate change ($P=1$) . Consider $\theta_1$ a measurement of temperature and $\theta_2$ precipitation ($\theta=(\theta_1, \theta_2)$). Let $\tau(\theta)$ be the adaptive actions that the community performs to affect their well being.  Then ${^aa_d}(\theta)$ shows how the well being is affected by this change in the adaptive action given a marginal change in the driver $\theta_d$. ${^ba_d}(\theta)$ represents the change of well being changes, due to adaptive behavioural changes with temperature or precipitation. 
If ${^ba_2}(\theta) < {^ba_1}(\theta)$, then adaptive capacity builds up quicker for temperature than for precipitation. Finally, ${^ca_{md}}(\theta)$ shows how a marginal change in the driver affects these adaptive actions, i.e, the optimal change in action given by a marginal change in temperature or precipitation. If ${^ca_{m2}}(\theta)>{^ca_{m1}}(\theta)$ then adaptation to precipitation requires a larger change in behaviour with respect to action $m$ in order to adapt to precipitation than temperature. If there are multiple actions that can be adjusted to changing drivers these relationships may vary per action.

\subsection{Total Impact (TI)}

The Total Impact (TI) combines exposure, sensitivity, and endogenous adaption. It is the overall change of the system property once exposed to the change in drivers and endogenous adaptation occurs.  TI measures changes in drivers on the system property. It is equal to sensitivity plus adaptive capacity. The latter is always positive. If sensitivity reduces the outcome of the system property, adaptive capacity counteracts this effect, otherwise enhances it.

\subsubsection*{Absolute}
The system property is evaluated at the initial value of the drivers with no adaptation $\psi({^0\theta}, \tau({^0\theta}))$, and at the new values with adaptation $\psi({\theta}, \tau({\theta}))$. The difference between both is defined as TI (Eq. \eqref{eq:vulnerability}).

\begin{align}
TI(\theta,{^0\theta}) &= S({\theta},{^0\theta}) + {^aA}({\tau(\theta},{^0\theta})) \nonumber \\
&= \psi(\theta, \tau(\theta)) - \psi({^0\theta}, \tau({^0\theta})) \nonumber \\
&= (\psi_1({\theta}, \tau({\theta})) - \psi_1({^0\theta}, \tau({^0\theta})),\dots, \psi_P({\theta}, \tau({\theta})) - \psi_P({^0\theta}, \tau({^0\theta})) ) 
\label{eq:vulnerability}
\end{align}

\subsubsection*{Marginal}

The marginal TI is the Jacobian of Eq. \ref{eq:vulnerability}. The entries of the Jacobian are defined by Eq. \ref{eq:vulneraMarg}. These show the change in the system property $p$ with an optimal adaptation $\tau(\theta)$, given a marginal increase in driver $\theta_d$. The marginal TI evaluated at the zero exposure levels of the drivers ${^0\theta}$ will be equal to the marginal sensitivity, as the marginal adaptive capacity is zero at that point.

\begin{align}
TI_{pd}(\theta) &=  \frac{\partial TI_p(\theta,{^0\theta})}{\partial \, \theta_d} = \frac{\partial \psi_p(\theta, \tau(\theta))}{\partial \, \theta_d}
\label{eq:vulneraMarg} \\
 &= s_{pd}(\theta)+{^aa}_{pd}(\theta) \nonumber 
\end{align}

\section{Case Study: North Sea flatfish fishery}
\label{sec:caseStudy}

We apply the framework to fishers' profitability in the North Sea flatfish fishery. The EU derives 32\% of the total landings from the North Sea and the Eastern Arctic, accounting for the highest total landed value in Europe \citep{STECF2019}. Historically, the most harvested species in this region by value are Atlantic cod, Atlantic mackerel, and Atlantic herring \citep{STECF2019}. However, a variety of other species such as European plaice, Common sole, and Common shrimp account for one third of the economic value generated in the North Sea. Fishing pressure caused shifts in the ecosystem composition historically and further shifts are expected due to climate change. This region is identified as one of the 20 hot-spots of climate change globally \citep{Pinnegar2016}.
\citet{Quante2016} show projections regarding increased sea level, ocean acidification, ocean temperature, and a decrease in primary production. This causes migration of the species, affecting the availability of resources to local fishing fleets, and reducing the overall ‘carrying capacity’ of the stock \citep{Pinnegar2016}.

The North Sea flatfish fishery is a multi-species fishery catching plaice, sole, cod, and other flatfish. The economic importance of fisheries in the North Sea led to over-fishing of some flatfish species. In this paper we focus on European Plaice (\textit{Pleuronectes platessa}) and Common Sole (\textit{Solea solea}), because they are the two principal flatfish species targeted by European fisheries \citep{Etherton2015}. Sole grows up to a length of 30cm, and plaice up to 33cm \citep{Knijn1993}. These species have endured the consequences of climate change, over-fishing, and pollution \citep{Engelhard2011, Gattuso2018}.  

\begin{figure}[h]
 \begin{minipage}{\textwidth}
  \includegraphics[width=\linewidth]{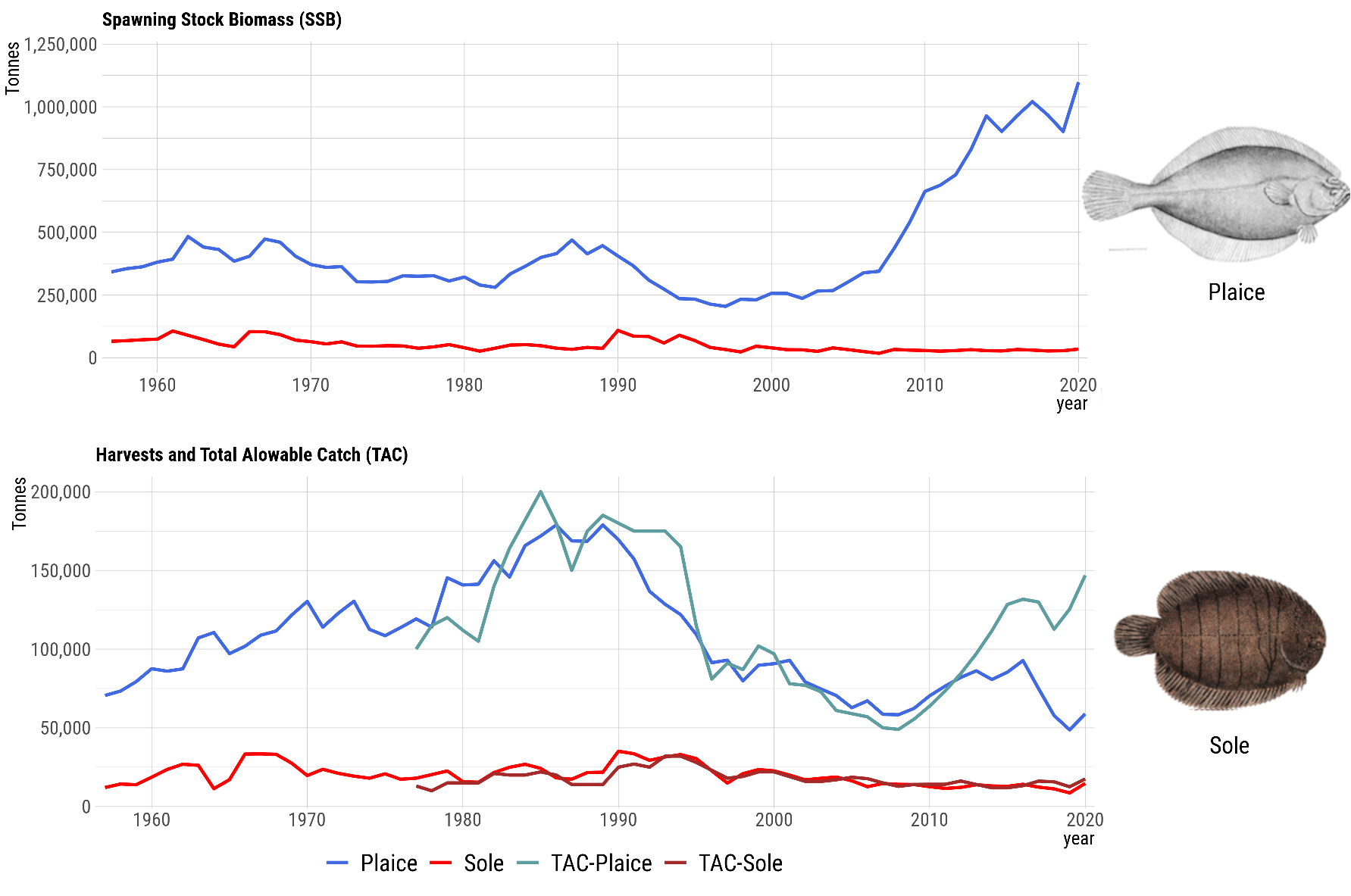}
  \caption[]{Spawning Stock Biomass (top), Harvests (Landings) and Total Allowable Catch (bottom) for plaice and sole between 1957-2020.} 
  \label{fig:ssb}
  \end{minipage}
\end{figure}

To promote the sustainability of the stock a policy was adopted regulating Total Allowable Catches (TACs), conservation areas, and mesh size  \citep{VanKeeken2007,Engelhard2011, EuropeanCommission2014}. TACs are in place since 1979 mostly restricting harvest of sole, while TACs for plaice have often been so large as to be non binding (Figure \ref{fig:ssb}) \citep{Daan1997}. During the second half of the 20th century, the TACs decreased for plaice, in line with a recommended reduction in fishery mortality \citep{Daan1997}. In 1989, to allow the plaice population to recover, a protected area, the ‘Plaice Box', is closed to trawling fisheries (an area on the Dutch and German coast). The Spawning Stock Biomass (SSB) for plaice decreases after this measure, attributed to a distribution shift caused by long term climate change and an increase in discards outside of the ‘Plaice Box' \citep{Engelhard2011,VanKeeken2007} (Figure \ref{fig:ssb}). The drop in the SSB for sole since 1990 was also caused by shifted distributions but strongly attributed to fishing pressure. The high price of sole makes it the preferred targeted fish compared to plaice \citep{Engelhard2011}, however, it is not possible to catch sole independently of plaice. In recent years the plaice stock (SSB) has recovered while sole shows a constant tendency \citep{ICES2015, ICES2019}

In the last decade, the average landings (harvests) of plaice by weight are approximately seven times larger than those of sole. However, because the price of sole is six times that of plaice, the two species' landings are roughly equal in value \citep{STECF2019}. In this region the price is controlled by companies in The Netherlands because it is the larger producer of European plaice in the world \citep{EUMOFA2013}. The main actors in this fishery are The Netherlands, Denmark, UK, Belgium, France, and Germany. Despite increases in costs net profits remain positive, except for the  Belgian and German fleets between 2010-2017 \citep{STECF2019}. 

In this paper we focus on three main aspects affecting this fishery. First, increasing technical measures affecting returns to effort. On 2023 the European commission called on members states to increase the monitoring and data collection of fishers to reduce the impact on the sea bed by bottom trawlers. By the end of 2024 member states are called to submit a national plan with the specific measures directed to data collection and monitoring programmes to improve observations and reporting of incidentally by-caught species \citep{EuropeanComission2023}. These measures could include the installation of cameras on board or additional requirements on the fishing measurement process. Such measures can cause less returns per unit of fishing effort in this fishery. 

Second, increasing regulations regarding the coverage of Marine Protected Areas (MPA) and Off-shore Windfarms (OWF) affecting the stock harvesting efficiency. The objectives of the EU 2030 Biodiversity Strategy is to protect 30\% of the European sea, and mobile bottom fishing in all MPA's by 2030. To achieve this objective the European commission calls the member states to create new MPAs and starting to adopt national measures by the end of March 2024. Offshore wind is also an increasing tendency, the European Commission estimates that by 2050 30\% of future global electricity demand could be supplied by offshore wind. Both, MPAs and OWF, reduce the fishers space available to fish and in the short term the stock available to fish, having effects on the stock harvesting efficiency.

Third, the ageing of the fishing population present an additional pressure on this fishery. \citep{STECFsocial2020} mentions that there is an inter-generational deficit which represent an important threat to the sustainability of this fishery. More than 60\% of the fishers are between 40-65 years old and only 22\% are between 25-39 years old. In 2019 this fishery experienced a sharp decrease (-18\%) in employment compared to 2018 \citep{STECF2021}. This could be the result of adaptation to simultaneous stressors, such as stocks moving towards another region, increasing fixed costs, shrink in active vessels, and reduced harvesting efficiency.

\subsection{Bio-Economic model}	
\label{sec:biomodel}

To apply the framework to our case study we use an existing bio-economic model \citep{Blanz2019}. In the context of an interconnected coupled ecological-economic system the bio-economic model is the most parsimonious product that incorporates the interconnections in a quantified way. We then calibrate the bio-economic model as it can serve as an intermediate complexity step between a fully conceptual model example and a pure data-based statistical analysis. Our framework, however, can be also used with another mathematical models or contexts.

In our application we modify the model described by \citep{Blanz2019} to embody the peculiarities of the North Sea flatfish fishery (See the detailed description of the model in \autoref{sec:appendix1}). We replace the logistic growth function, used to model stock change, with a Ricker-recruitment type growth function \citep{Ricker1975}. We also introduce weighting factors for each fish species in the household utility function to better reflect consumer preferences.  A feature of the model is the introduction of simultaneous multi-species harvesting, i.e., fisheries target one species but in doing so catch other species. In our case study, the fishers behaviour is market-driven. Fishers mostly target sole because of its higher price, but in doing so they also catch plaice \citep{Aarts2009}. The model includes parameters that account for these characteristics to resemble observations.

The bio-economic model has three elements: (i) The ecosystem component includes harvests and the stock change, represented by the species growth function for plaice and sole. The stock levels are the system's state variables. The system's stable and non-stable steady-states depend on the stock change which results from ecosystem growth and harvests.
(ii) The harvesting component includes an endogenous amount of fisheries firms comprising the fleets of two m\' etiers\footnote{M\' etier refers to a combination of vessel and gear type. In this paper we use a model with two species where the species sub-index \textit{i} takes the value of 1 for plaice and 2 for sole. Similarly the sub-index \textit{k} refers to the two fleets, where \textit{k=1} refers to the fleet targeting plaice and \textit{k=2} the fleet targeting sole.}. The first targets 
plaice and the second sole with imperfect selectivity. The harvesting function depends on effort and stock availability. Firms maximize profits, derived from harvests, prices, variable and fixed costs.
(iii) The household component consists of a representative household obtaining utility from fish consumption and manufactured goods. The household maximizes utility subject to a budget restriction and thereby determines the optimal quantities demanded and willingness to pay for each fish species.	

The model assumes market-clearing, all goods produced are consumed (Eq. \ref{eqn:marketclearing}). In the long run a competitive market with free entry and exit, firms compete such that prices and total costs are equal. This leads to the zero profit assumption described in Eq. (\ref{eqn:maxFirms}).  The size of the fleets is determined satisfying the zero profit assumption and the optimal effort choice by fishing firms.	There is no fishery rents since effort adjustment is much faster than fleet adjustment, so we consider fleet fixed when investigating effort.  Another assumption of the model is that processors set prices to take any rents irrespective of consumer demand and do not adjust prices. This assumption is based on qualitative information of reality because German fishers are price takers given the monopoly of the price established by companies from The Netherlands \citep{EUMOFA2013}.
The steady-states for stocks of each species in an open-access scenario with quotas are determined numerically. Our model resembles particular aspects of the North Sea flatfish fishery mainly catching plaice and sole. The model presents an abstraction of the multiple complexities embedded in this fishery, but still useful providing  insights regarding the vulnerabilities we analyse. Although the model includes by definition many assumptions, these are not required to apply the framework.

\subsection{Calibration of the model}

We calibrate the model to time series of stocks, harvests, and prices for the whole North Sea. For stocks and harvests, we use data on spawning stock biomass (SSB) and landings from 1957 to 2019 provided by the International Council for the Exploration of the Sea \citep{ICES2019, ICES2015}. We use price data from the European Market Observatory for Fisheries and Aquaculture Products (EUMOFA) database for the years from 2000 to 2020. The ecosystem component is calibrated independently of the economic parts using the observed stock growth and harvests. Within the model, harvests and consumer demand are calculated based on the stock levels of each period. To account for this the economic parameters of the model are calibrated to harvests and prices of each period simultaneously. A detailed description of the calibration method is provided in \autoref{sec:appendix2}.

\begin{figure} [h]
  \includegraphics[width=\linewidth]{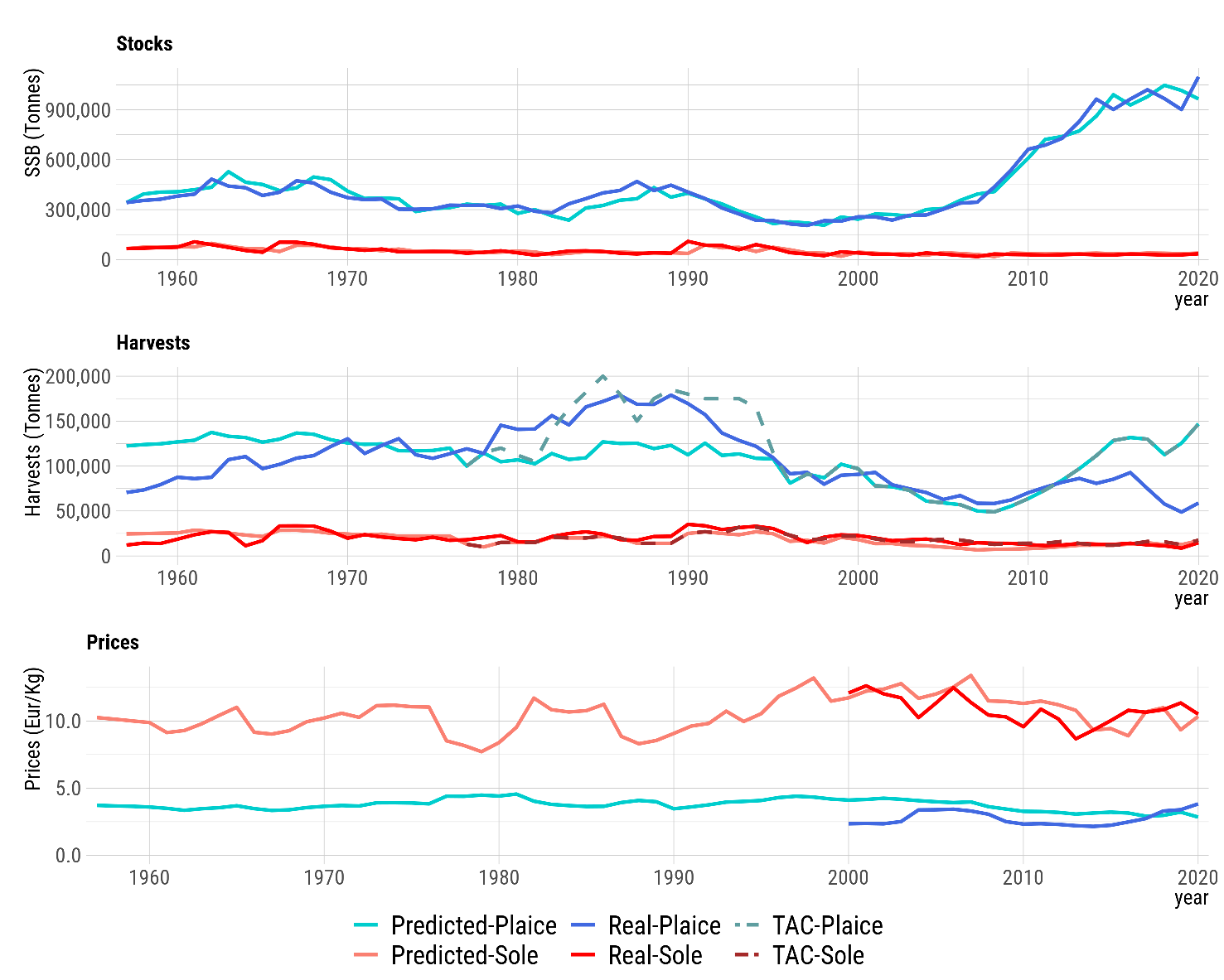} 
  \caption{The difference between the real data and predicted levels of stocks (top), harvests (centre), and prices (bottom). Predicted stock levels are the result of predicted growth, given the real data in the previous period.  The shown predicted levels of harvests and prices are based on the real stock levels of each period.   
Theil Inequality Coefficient: Stocks: $\text{Plaice}=0.049$, $\text{Sole}=0.1507$ Harvest: $\text{Plaice}=0.1506$, $\text{Sole}=0.1389$ Prices: $\text{Plaice}=0.1615$, $\text{Sole}=0.0516$ 
  }
  \label{fig:realPredict}
\end{figure}

Tables \ref{Tab:parmsMeaning} and \ref{Tab:steadyMeaning} present the calibrated and output values of the model elements in steady-sate. Figure \ref{fig:realPredict} shows the output of the calibration for SSB (stock), harvest, and prices. The predicted values for SSB resemble the real tendency of the stocks during the last forty-five years. The harvest predictions of plaice before the TAC was introduced are higher than the real time-series. This is because the modelled fleet adjusts automatically to the new levels of stocks and prices, while in reality, the enter-exit movement of the firms occurs over a longer time frame. The predicted values of plaice show a decreasing price from 1982 to 1986 followed by a decreasing harvest. After the introduction of the plaice box in 1989 the plaice price increased together with plaice harvest until the TAC becomes binding in 1995. The predicted values of sole harvest follow the binding TAC. Since 1987 the predicted sole price starts increasing followed by a slight increase in harvest until 1999 when the TAC decreases again. For the last ten years, the predicted sole harvest and prices resemble the real values. However, the predicted plaice harvest follows the path of the TAC because the real fishing capacity do not keep up with the TAC and the plaice industry do not profit much from it since the plaice price is low.

\begin{table}[!htbp]
\centering
\scriptsize \addtolength{\tabcolsep}{-2pt}
\renewcommand{\arraystretch}{1.5}
\scalebox{0.8}{
\begin{tabular}{|c|M{1.7cm}|M{7cm}|M{3cm}|M{3.1cm}|}
\hline
\rowcolor{gray!10}
\textbf{Symbol} & \textbf{Value}& \textbf{Description}& \multicolumn{2}{|c|}{\textbf{Exposure}}\\
\hline
\multicolumn{3}{|l|}{\textbf{Ecosystem drivers}}&Absolute (Min, Max)&Percentage (Min, Max)\\
\hline
\multirow{2}{*}{$x_i$}& $x_1 = 148.589$ & \multirow{2}{*}{\parbox{5.7cm}{
																				\centering Steady-state 
																				output for stocks of plaice 
																				and sole in tonnes.}}        &  & \\
				
                      & $x_2 = 85.936$ & 											      &  &\\ 
\hline
\multicolumn{5}{|l|}{\textbf{Harvesting drivers}}\\
\hline
$\epsilon$ & 0.5 & Returns to effort. A higher value of $\epsilon$ refers to less returns per unit of effort. & 0.48, 0.52 &$~-3.09\%,+4.26\%$\\
\hline
\multirow{2}{*}{$\chi_i$}& $\chi_1 = 0.308$ & \multirow{2}{*}{\parbox{7cm}{
																				\centering Stock harvesting 
																				efficiency of the species $i$. 
																				Represents the ability to catch a
																				 species depending on stocks 
																				 availability (catchability).}}  & $\chi_1: 0.093, 0.607$ &$\chi_1: -69,7\%,+96.8\%$ \\
				
                         & $\chi_2 = 0.308$ & 											      & $\chi_2: 0.230, 0.549$ &$\chi_2: -25,2\%, +78.0\%$\\ 
\hline
\multirow{4}{*}{$\nu_{ik}^\ddagger$}& $\nu_{11}=1.00$ & \multirow{4}{*}{\parbox{7cm}{
																				\centering M\' etier specific harvesting 
																				efficiency  ($\nu_{ik}$) of the species 
																				$i$ targeted with the m\'etier $k$.}}& & \\
				
                         						& $\nu_{12}=0.75$    & 									& &\\ 
                         						& $\nu_{21}=0.00$    & 									& &\\
                         						& $\nu_{22}=0.25$    & 									& &\\                         \hline
\multicolumn{5}{|l|}{\textbf{Market drivers}}\\
\hline
\multirow{2}{*}{$p_i$}& $p_1 = 5.6$ & \multirow{2}{*}{\parbox{5.7cm}{
																				\centering Market prices in 
																				(Euros/Kg) for plaice $p_1$
																				 and sole $p_2$ in 
																				 steady-state. }}        & &\\
				
                      & $p_2 = 6.6$ & 											       & &\\ 
\hline
$\omega^\ddagger$& 1 & Wages. The model wage is normalized to one, and households receive a unit to spend in either other goods or fish.& 0.65,1.37 &$~-35\%,+37\%$\\
\hline
 $\phi$ & 1.0 x$10^{-8}$ & Fixed costs of harvesting firms. Costs of owning the harvesting vessel and equipment independent of use.&  &\\
\hline
\multicolumn{5}{|l|}{\textbf{Household preferences drivers}}\\
\hline
$\alpha$ & 6.77 x $10^{-5}$ & Relative importance of fish consumption for households.  
&& \\
\hline
\multirow{2}{*}{$\beta_i$}& $\beta_1 = 2.69$ & \multirow{2}{*}{\parbox{5.7cm}{
																				\centering Weight of the species $i$
																				 in the  household utility 
																				 function. }}        						 & &\\
				
		                      & $\beta_2 = 4.14$ & 											       & &\\ 
\hline
$\eta$ & 1.10 & Elasticity of demand for fish consumption.  & & \\
\hline
$\sigma$ & 2.01  & Substitution elasticity between plaice and sole.  && \\
\hline
\end{tabular}
}
\caption{Calibration results for each parameter, and steady-state values for prices, and stocks.  $^\ddagger$These parameters are not included in the calibration and are taken from the theoretical results in \citet{Blanz2019}.}

\label{Tab:parmsMeaning}
\end{table}

\begin{table}[!h]
\centering
\scriptsize \addtolength{\tabcolsep}{-2pt}
\renewcommand{\arraystretch}{1.5}
\scalebox{0.8}{
\begin{tabular}{|c|M{2cm}|M{11cm}|}
\hline
\rowcolor{gray!10}
\textbf{Symbol} & \textbf{Value}& \textbf{Description}\\
\hline
\multicolumn{3}{|l|}{\textbf{Steady state values}}\\
\hline
$n_i$ & $n_1=383$, $n_2=2315$ & Optimal number of firms for each species.    \\
\hline
$h_{ik}$ & $h_{11}=13.752$, $h_{12}=62.317$, $h_{21}=0.00$, $h_{22}=17.545$ & Optimal harvests ($h_{ik}$) of species $i$ per metiér $k$ in tonnes.  The fleet targeting plaice ($k=1$), only catches plaice.  \\
\hline
$e_k^*$ & $e_1= 1.0$x$10^{-8}$, $e_2= 1.0$x$10^{-8}$ & Optimal effort in steady-state for the metiér $k$. This is the effort that results from the zero profit condition and profit maximization(Eq. \ref{eqn:optEffortZPC}). 
  \\
\hline
\multicolumn{3}{|l|}{\textbf{Scaling parameters}}\\
\hline
$\kappa$ & $533.459,8$   & Scaling parameter for stocks.  The real values of SSB and landings were divided by this parameter to scale to model values.  \\
\hline
$wScale$	 & $10.052.180$x$10^6$   & Scaling parameter for the income of the economy.  This value correspond to the whole economy GDP of the North Sea countries for the year 2015.  \\
\hline
\end{tabular}
}
\caption{Calibration results for steady-state values of firms, harvests and effort. $\kappa$ and $wScale$ are used to scale the real data to model values. }
\label{Tab:steadyMeaning}
\end{table}

\newpage
\subsection{Application of the analytical framework to the bio-economic model}

The framework presented above enables us to find the adaptations of many system properties to many drivers. Hence, the main question to answer before proceeding with the case study is the adaptation \textit{of} what \textit{to} what?. We select fishers' economic viability to answer the first ``what'' as the most critical aspect in this sector \citep{Schuhbauer2016}. For the second ``what'' we assess drivers
derived from changes in policies affecting the harvesting process ($\theta$). After identifying the best adaptation of fisher's profits drivers considered we identify the effect of this adaptation on the quantities produced and the utility in the economy.

In our application, we replace $\psi(\theta)$ by $\pi(\theta)$, which corresponds to the fishers' profits. There are two fishers' m\' etiers ($k \in \lbrace1,2\rbrace$) that fish two species ($i \in \lbrace1,2\rbrace$). We evaluate profits of two metiers, hence $\pi(\theta)=(\pi_1(\theta), \pi_2(\theta))$. Profits are a function of the set of drivers ($\theta$) and depend on harvests ($h_{ik}$) of species $i$ with m\' etier $k$, prices ($p_i$) of species $i$, effort ($e_k$) of the m\' etier $k$, wages ($\omega$), and fixed costs ($\phi$) (Eq. \ref{eq:profitwithEfortBefore}). We analyse profits before the ‘zero profit condition' holds to allow profits to deviate from zero (Eq. \ref{eqn:maxFirms}). We investigate the short term effects on individual fishing companies. Market forces will drive profits to zero by entries and exits of firms in the long term. Our analysis precedes these adjustments. I.e., if profits are likely to decrease/increase due to changes in a driver, this forms the incentives for firms to enter or exit the market in the longer term. In our case study we replace the adaptation mechanism  $\tau^{*}_m(\theta)$ by effort $e_k^{**}(\theta)$ \footnote{In our application we evaluate two properties $K=2$ and use two adaptation behaviours for each property that correspond to the effort of each m\' etier ($M=2$). Because each property corresponds to an adaptation behaviour we use the same index $k$ for both. Our framework allows multiple adaptation mechanisms for one system property, but in this application we use only one.}. This effort is the best adaptation a fisher can achieve to maximize fishers profits, before the zero profit condition holds. The modelled fisher adapts to changed conditions by modifying fishing effort  (Eq. \ref{eq:adaptiveEffort}). We name ($e_k^{**}$) the adaptive effort to distinguish from the equilibrium effort ($e_k^*(\theta)$) which is derived once the zero profit condition holds (Eq. \ref{eqn:optEffortZPC}).

\begin{equation}
 \pi_{k}(\theta) = \sum\limits_{i=1}^{\bar{i}}  h_{ik}(e_{k}^{**}, x_{i})p_{i} - \omega e_{k}^{**} - \phi   
 \label{eq:profitwithEfortBefore}
\end{equation} 
In Eq. \ref{eq:profitwithEfortBefore} harvest ($h_{ik}$) and adaptive effort ($e_k^{**}$) are defined in Eq. (\ref{eqn:harvestspermetier}, \ref{eq:adaptiveEffort}) where $x_i$ is the available stock, $\nu_{ik}$ is the m\' etier harvesting efficiency, $\chi_i$ the stock harvesting efficiency, and $\epsilon$ the returns to effort.

 \begin{equation}
  h_{ik}(e_k^{**}, x_{i}) = \nu_{ik} (e_k^{**})^\epsilon x_{i}^{\chi_i} 
    \label{eqn:harvestspermetier}  
 \end{equation}

\begin{equation}
e_{k}^{**} (\theta)= \left(   \frac{\epsilon}{\omega} \sum\limits_{i=1}^{\bar{i}} \nu_{ik} x_{i}^{\chi_i} p_{i} ) \right) ^\frac{1}{1 - \epsilon} 
 \label{eq:adaptiveEffort}
\end{equation}

\subsubsection{Drivers}

We evaluate three drivers according to the factors that we consider to be the most critical in this fishery. First, we evaluate returns to effort ($\epsilon$). We argue that regulations changing the monitoring of fishing and requirements on data collection, as mentioned in section \ref{sec:caseStudy}, affect the returns to effort by changing fishers' working conditions. Second, we assess changes in stock harvesting efficiency for each species ($\chi_i$). $\chi_i$ is affected by regulations changing the space available to fish such as, MPAs and OWF. In the short term fishers experience less available fish affected by the harvesting efficiency. Third, we assess changes in wages ($\omega$). This fishery is ageing  representing a threat to its sustainability, we identify the effect of changes in wages on the optimal adaptation through effort. We consider that an increase in wages could attract the new generations such that it can keep the fishery active.

\subsubsection{Exposures}

Exposure is defined as changes in values for each element of $\theta$ (Eq. \ref{eq:exposure1}). The magnitude of exposure for each driver is based on historical variations of harvests, stocks, prices, wages, and fixed costs observed in the data. We use the harvest variations to identify the exposure limits for the m\'etier harvesting efficiency ($\nu_{ik}$), returns to effort ($\epsilon$), and stock harvesting efficiency ($\chi_i$). Using Eq. (\ref{eqn:harvestspermetier}) we obtain the maximum and minimum intervals of each driver that result in the same harvesting range. Exposure levels of stocks ($x_i$), prices ($p_i$), wages ($\omega$) and fixed costs ($\phi$) are taken from maximum variations in the data \footnote{The sample of maximum and minimum variations are within the 95\% of confidence intervals.}.  We use the values of the North Sea countries with the maximum deviations as a reference for exposures. For the household drivers ($\sigma, \alpha,\beta_i$, and $\eta$), the boundaries match the upper and lower bounds reflected in the harvest intervals using Eq. (\ref{eq:demandedQuantity1}). The selected exposures for each driver are described in the last column of Table \ref{Tab:parmsMeaning}. They are interpreted relative to the steady-state values, i.e., the status-quo of the system from which the vulnerabilities are analysed.

\subsubsection{Sensitivities}

We characterize the sensitivities of fishers' profits to drivers from the ecological, harvesting, market, and household components (Table \ref{Tab:parmsMeaning}).  Sensitivities are described using Eq. (\ref{eq:sensitivity1}). We analyse individual sensitivities of profits for each driver, holding other drivers constant. An example of the absolute sensitivity of profits to changes in stock harvesting efficiency ($\chi_i$) is Eq. (\eqref{eq:sensitivityDisprofits1}). 
$\chi_i$ is the new level of exposure and $^0\chi_i$ the original value, keeping stock and prices constant at steady-state levels. We apply the same exercise for ecosystem, harvesting, and market drivers. For stock changes ($x_i$) prices are constant, and for changes in prices ($p_i$), stock is constant. 

\begin{equation}
S_k(\theta, {^0\theta}) = {\pi_k}(\theta, e_k^{**}(^0\theta)) - \pi_k (^0\theta, e_k^{**}(^0\theta))
\label{eq:sensitivityDisprofits1}
\end{equation}

\begin{align}
\theta &= (0,\dots,\chi_i,\dots,0) \nonumber\\
^0\theta &= (0,\dots,^0\chi_i,\dots,0) \nonumber
\end{align}

We use Eq. (\ref{eqn:calibOptFun}) to find the sensitivities of profits to household drivers. This equation is derived from the household optimization procedure and represents the demand for one good given the consumption of the other. 
Profits are affected by household drivers through the demand side, i.e., changes in these drivers affect the quantity demanded, then the effort is adjusted to this new quantity and later profits change. In our analysis the market clearing condition holds, i.e., harvest is equal to the demanded quantities (Eq. \ref{eqn:marketclearing}). The optimal prices ($p_i$) and demand ($q_{-i}$) at steady-state are held constant for the analysis not to muddle effects.  We use Eq. (\ref{eq:demandedQuantity1}) to derive sensitivities from the household component.

\begin{equation}
  H_2 = q_2 = \left(  \left(\frac{p_1}{\alpha \beta_1} (\beta_1 q_1)^{\frac{1}{\sigma}} \right)^{\frac{\eta(\sigma-1)}{\eta -\sigma}}  - (\beta_1 q_1)^{\frac{\sigma - 1}{\sigma}}  \right)^{\frac{\sigma}{\sigma-1}} (\beta_2)^{-1}
  \label{eq:demandedQuantity1} 
\end{equation}

The sensitivities of profits for the m\' etier $k$ to a marginal change in the driver $d$ are defined by Eq. (\ref{eq:sensitivityConReal}). Profits are evaluated with the adaptive effort  ($e_k^{**}$) embedded, not the equilibrium effort ($e_k^{*}$), hence this derivative is different than zero.

\begin{equation}
s_{kd}(\theta) = \frac{\partial \pi_k(\theta, e_k^{**}(^0\theta))}{\partial \theta_d} 
\label{eq:sensitivityConReal}
\end{equation}

\subsubsection{Adaption}

We determine the adaptation by evaluating the difference in profits in two cases. We elicit profits when fishers first experience the change in the driver ${\pi_{k}}(\theta, e_k^{**}(^0\theta)) $, without yet modifying their effort. Then, we identify profits after adaptation ${\pi_{k}}(\theta, e_k^{**}(\theta))$,
once the effort is adjusted to the new level of the driver $e_k^{**}(\theta)$. The difference in profits between these two values yields the absolute benefit of adaptation per m\' etier $k$ for each driver ($\theta_d$). We assess individual adaptive capacities for each driver $\theta_d$ holding others constant following the same procedure as with the sensitivities. 

\begin{equation}
{^aA}_k(\theta, {^0\theta}) = {\pi_k}(\theta, e_k^{**}(\theta)) - {\pi_k}(\theta, e_k^{**}(^0\theta)) 
\label{eq:adaptiveProfits}
\end{equation}

The marginal benefit of adaptation for fishers' profits using the adaptation effort with m\' etier $k$ to the driver $d$ are in Eq. (\ref{eq:adaptiveConReal1}. \ref{eq:adaptiveConReal2}). Eq. (\ref{eq:adaptiveProfits}) shows the optimal change in adaptive behaviour due to a change in the driver. Eq. (\ref{eq:adaptiveProfits},\ref{eq:adaptiveConReal1}) evaluated at steady state ($^0\theta$) are zero. In our framework adaptation is always positive, i.e., it is increasing with any deviation from zero exposure, consequently the derivative is zero at this point.

\begin{equation}
{^aa}_{kd}(\theta) = \frac{A_k(\theta, {^0\theta})}{\partial \, \theta_d} 
\label{eq:adaptiveConReal1}
\end{equation}

\begin{equation}
{^ba}_{kd}(\theta) = \frac{\partial^2 {A_k(\theta, {^0\theta})}}{\partial^2 \, {\theta_d^2}} 
\label{eq:adaptiveConReal2}
\end{equation}

\begin{equation}
{^ca}_{kd}(\theta) = \frac{\partial e_k^{**}(\theta)}{\partial \, \theta_d} 
\label{eq:adaptiveConReal}
\end{equation}

\subsubsection{Total Impacts}

We use Eq. \ref{eq:vulnerability} to derive the TIs of profits to multiple drivers. TIs are determined as the overall difference in profits at the initial level of the driver ($^0\theta_d$) and at the new level ($\theta_d$). Profits at the initial level of the driver and without adaptation yield: $\pi_{k}(^0\theta_d, e_k^{**}(^0\theta_d)) $. Profits at the new level of the driver and with adaptation included yield: ${\pi_{k}}(\theta_d, e_k^{**}(\theta_d))$. We assess the TI using Eq. (\ref{eq:vulnerabilityProfits}), for each driver independently.  

\begin{equation}
TI_k(\theta,{^0\theta}) = {\pi_k}(\theta, e_k^{**}(\theta)) - \pi_k(^0\theta, e_k^{**}(^0\theta)) 
\label{eq:vulnerabilityProfits}
\end{equation}

The marginal TIs contemplate the derivative of profits once there is an optimal adaptation to the change in the driver (Eq. \ref{eq:vulneraMargPi}). 

\begin{equation}
TI_{kd}(\theta_d) = \frac{\partial TI_k(\theta,{^0\theta})}{\partial \, \theta_d} 
\label{eq:vulneraMargPi}
\end{equation}

After identifying how fishers adapt to maximize profits we identify the effect of this adaptation on the fish quantities available in the market and the utility of the economy.

\section{Results}

\begin{figure}[h]
  \includegraphics[width=\linewidth]{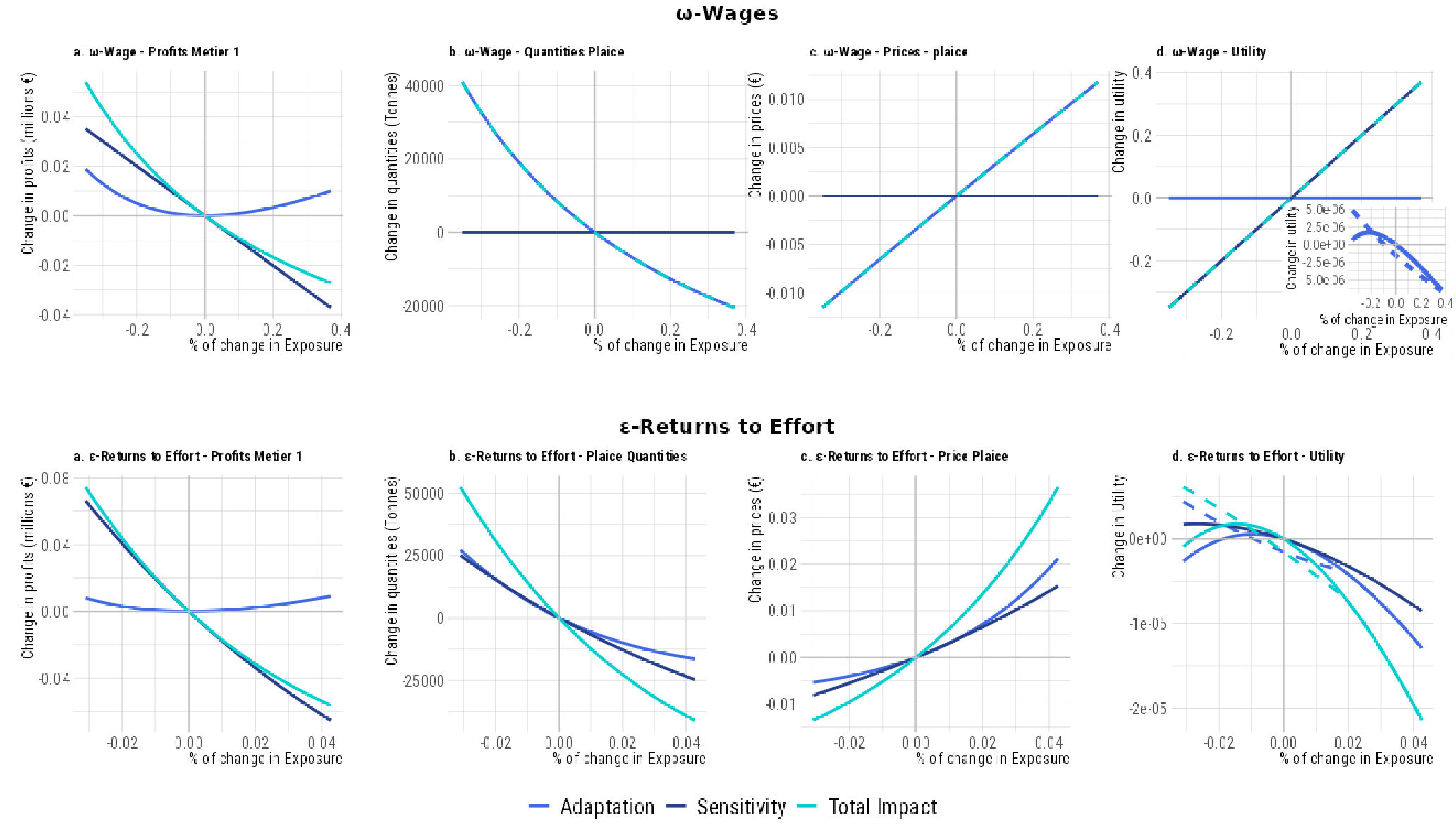}
  \caption {Changes in profits, quantities, prices and utility due to changes in wages.}
  \label{fig:parmsImpact}
\end{figure}

In our case study we investigated the sensitivity, adaption, and total impact (TI) \emph{of} fishing profits, utility and quantities \emph{to} changes in four drivers. The TIs, adaptations, and sensitivities of profits to drivers are presented in figure \ref{fig:parmsImpact} for wages ($\omega$) an returns to effort ($\epsilon$). The figures corresponding to the harvesting efficiency ($\chi_i$) are in the appendix \ref{fig:chiAnalysis}. The horizontal axes represent the magnitude of exposure for each driver ($\theta_d$) within the levels established in Table \ref{Tab:parmsMeaning}. The change on the vertical axes is calculated relative to steady-state. We perform the analysis from the steady-state to facilitate interpretation, however, within framework any other reference can be used. Profits are scaled relative to the household income ($\omega$).  As exposures are relative to the starting value, the initial level of exposure, adaptation, sensitivity, and TI is zero.

The first row of figure \ref{fig:parmsImpact} shows the impact of wages on profits of the m\'{e}tier 1, quantities and utility. The sensitivity shows that increasing wages increases the costs and hence diminishes profits. Fishers adapt by decreasing their effort to reduce costs and counteract the effect of sensitivity. The total impact shows the impact of wages in profits once fishers adapt. The quantities do not change with wages (\ref{eqn:harvestspermetier}) when the change in the driver is experienced. When fishers adapt the quantities change with effort. The prices show the willingness to pay of household given the quantities available in the market following equation \ref{eqn:willignestopay}. 

"The figure in the first row labeled 'd.' in \ref{fig:parmsImpact} displays the changes in utility. The immediate shock of wages affects utility linearly (see equations \ref{eqn:utility} and \ref{eqn:wages}). Once fishers adapt through effort, the quantities available in the market change. However, this change has a low effect on utility compared to the sensitivity. The small plot on the utility graph illustrates the effect of fishers' adaptation on changes in utility. An increase in wages decreases the quantity available in the market, and hence utility decreases. A decrease in wages increases the effort and, consequently, the quantities available in the market. The reduction in wages leaves households with less money available to buy fish or the numeraire good. When wages decrease more than 20\%, households are supposed to buy the new quantities offered; however, when buying more than demanded, their utility decreases.

In our analysis, we only focus on adaptation through effort, keeping all other variables constant; i.e., households consume the new quantities offered by the fishers without household adaptation. The dotted line shows the changes in utility once households adapt to the new quantities of fish, a case in which they reduce fish consumption, increasing utility. Overall, adaptation increases fishers' profits; however, the effect of adaptation on utility is almost null. Increasing wages increases utility, so that households mainly increase their consumption of the numeraire good. This is due to the low level of preferences for fish consumption ($\alpha$).

The second row of figure \ref{fig:parmsImpact} presents the changes in profits, quantities, prices, and utility due to a change in returns to effort ($\epsilon$). A policy reducing harvesting efficiency corresponds to an increase in $\epsilon$. A higher $\epsilon$ decreases effort, followed by a reduction in the harvest, i.e., the quantities available in the market. It causes an increase in prices and decreases utility. To adapt, fishers reduce effort, which reduces their costs, counteracting the sensitivity and increasing profits. Note that the effect of adaptation on profits is always positive. The reduction in effort due to the increase in $\epsilon$ reduces fish quantities, increasing prices and decreasing utility.

\begin{figure}[h]
  \includegraphics[width=\linewidth]{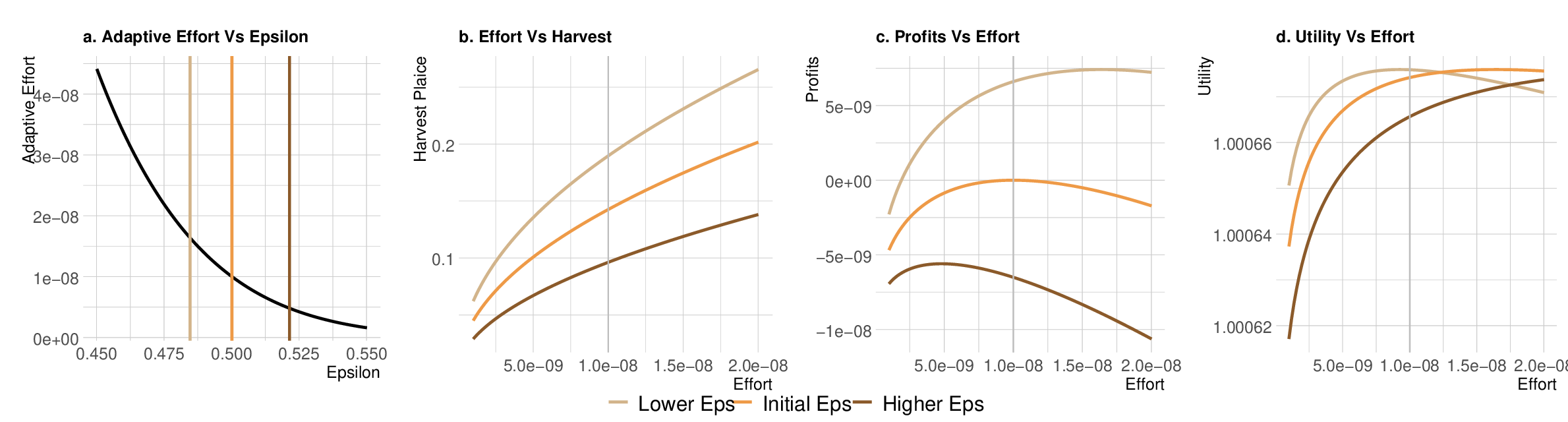}
  \caption {Changes in effort, harvest, profits and utility for initial, higher and lower values of $\epsilon$}
  \label{fig:epsAnalysis}
\end{figure}	

A decrease in $\epsilon$ increases effort, effective effort ($e^\epsilon$) and hence harvest (see Figure \ref{fig:epsAnalysis}). Figure \ref{fig:parmsImpact}.d at the second row shows that utility increases for a decrease in epsilon. However, when there is a large decrease in $\epsilon$ the utility decreases. This is because in this setting $\epsilon$ changes ceteris paribus, i.e, households consume all produced by the market. The increase in quantities produced by the fishers reach a point where households consume more quantities than what they demand and utility is reduced. The dashed lines show the changes in utility when households adapt to the new quantities offered by the market and the total impact after this adaptation. Lastly, changes in the availability of stocks modify $\chi_i$. A higher $\chi$ shows a lower availability of stocks for fishers. The analysis for $\chi_i$ resembles the reasoning of $\epsilon$ (see fig. \ref{fig:chiAnalysis}.d).

\begin{figure}[h]
  \includegraphics[width=\linewidth]{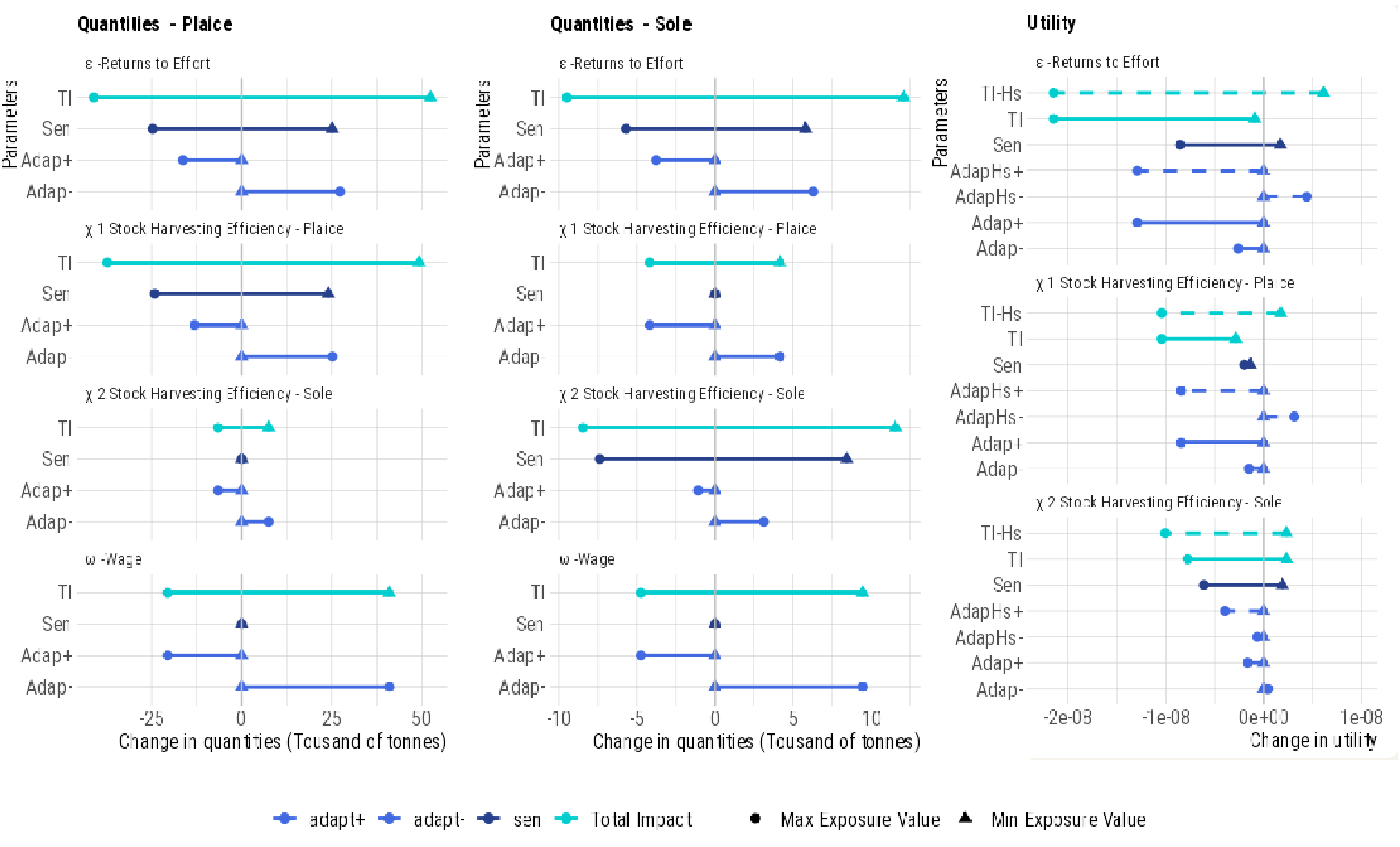}
  \caption {Absolute changes in quantities and utility due to changes in drivers.}
  \label{fig:absChangesQuanUtil}
\end{figure}

Figure \ref{fig:absChangesQuanUtil} shows the comparison of the effect of the considered drivers on quantities and utility for the extremes of the established exposure. The horizontal axis displays the minimum and maximum vertical values of profits presented in figure \ref{fig:parmsImpact}. 'Adapt+' represents the effect of adaptation on changes in profits when exposure increases, and 'Adapt-' when exposure decreases. The symbols represent the maximum and minimum values of each driver. Dotted lines represent the adaptation and total impacts in cases of household adaptation (Hs). Absolute changes in profits are presented in \ref{fig:overallProfits}.

The driver with the largest impact on quantities and utility is returns to effort. For changes in $\epsilon$ adaptation has a high effect on counteracting the sensitivity, and the effect in utility is low, specifically when it decreases. Changes in $\chi_1$  mostly affects the quantities of plaice and $\chi_2$ the sole quantities. $\chi_1$ affects more the quantities of sole offered than $\chi_2$ affecting the quantities of plaice. Decrease in wages have a higher impact in the quantities of plaice than sole. This is mainly because in steady state sole is restricted by quota and plaice is not. Wages have the highest impact in the utility as shown in \ref{fig:parmsImpact}\footnote{The effect of wages in utility is not presented in figure \ref{fig:absChangesQuanUtil} to maintain proportions. However figure \ref{fig:parmsImpact} shows the total impact on utility.}.

\begin{figure}[h]
  \includegraphics[width=\linewidth]{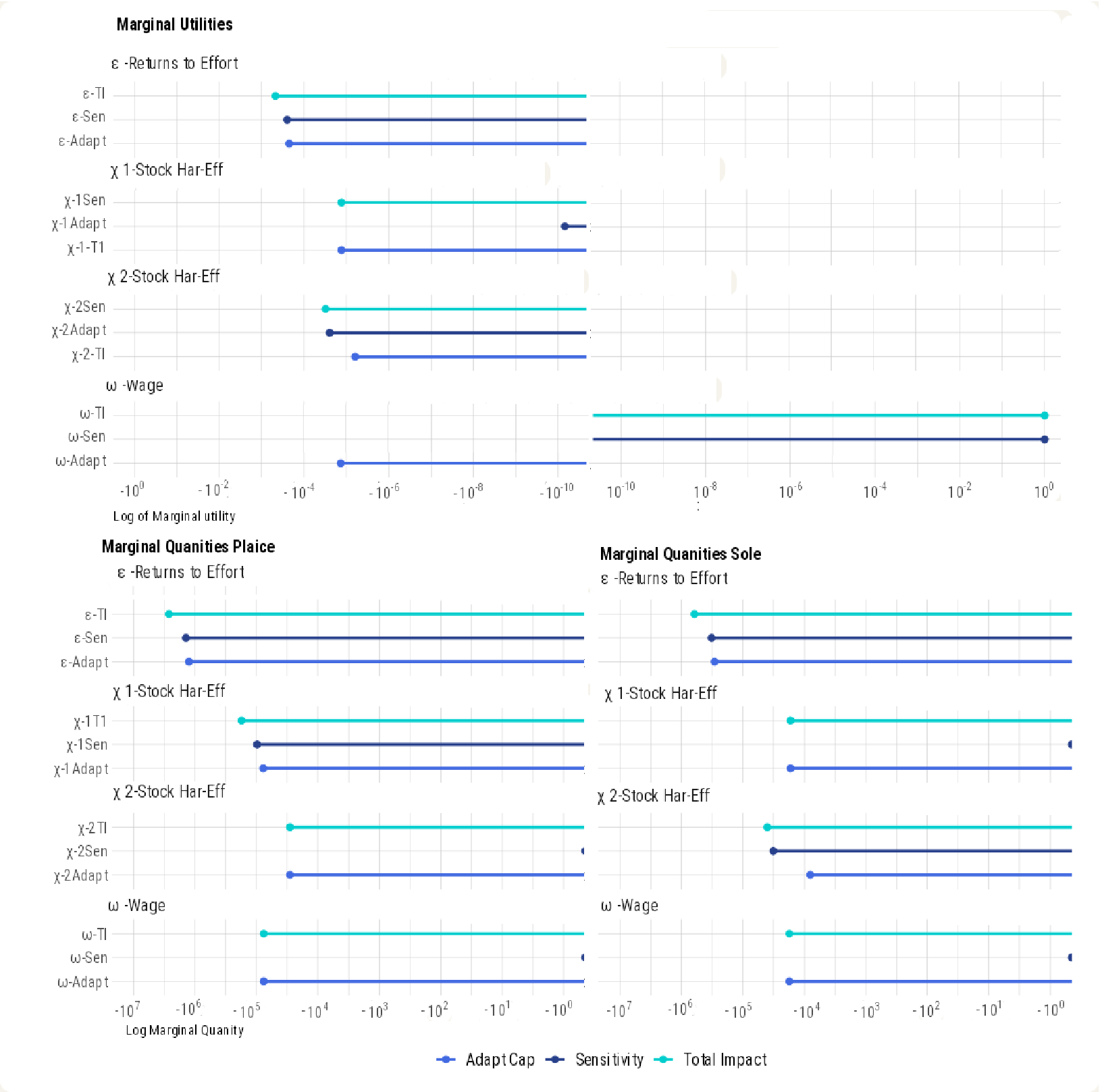}
  \caption {Marginal changes in quantities and utility due to a change in drivers.}
  \label{fig:marginalsQuanUtil}
\end{figure}	

Figure \ref{fig:marginalsQuanUtil} shows the marginal changes in adaption, sensitivities and total impacts for quantities and utility given changes in drivers. They are the result of the equation \ref{eq:adaptiveDis1} for quantities and utility with the drivers considered. Marginal increases in $\epsilon$ decreases the quantities of plaice more than sole, this is because in steady state the quantities harvested of plaice are more than four times those harvested for sole (The metier harvesting efficiency is higher for sole than plaice, see table \ref{Tab:steadyMeaning}). Marginal changes in the stock harvesting efficiency affect quantities of plaice more than sole. Changes in quantities due to marginal increase in wages ($\omega$) only changes through effort, hence the sensitivity is zero.   

Utility is mostly affected by marginal increases in wages. The marginal adaptation of fishers' effort to wages affects quantities more than utility. Marginal changes in $\chi_2$ affect the sensitivity of utility to a higher proportion than changes in $\chi_1$. This is because households value sole more than plaice (the elasticity of substitution ($\sigma$) between plaice and sole is 2). When fishers adjust the effort to a marginal increase in $\chi_i$, the effort is reduced because there is less available fish to harvest. Hence, the marginal quantities are reduced, leading to a decrease in marginal utility.


\begin{figure}[h]
  \includegraphics[width=\linewidth]{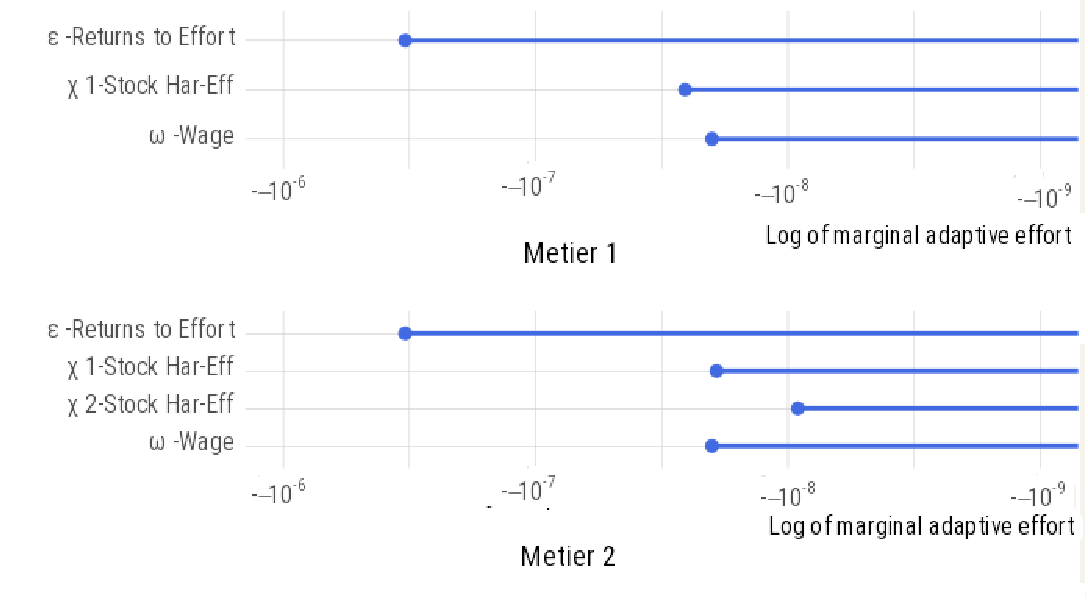}
  \caption {Marginal adaptive effort for the drivers considered in Metier 1 and 2.}
  \label{fig:marginalAdaptiveEffort}
\end{figure}	

The marginal adaptations ${^aa}_{kd}(\theta)$ in steady state $^0\theta$ are zero for all drivers (See Fig. \ref{fig:parmsImpact}. Hence, we present ${^ca}_{kd}(\theta)$ in figure \ref{fig:marginalAdaptiveEffort}. 
 ${^ca}_{kd}(\theta)$ shows the change in adaptive effort given by a marginal change in the driver (Eq. \ref{eq:adaptiveConReal}). The adaptive effort is mostly affected by returns to effort ($\epsilon$) followed by
stock harvesting efficiency ($\chi_1$) and wages ($\omega$).

The absolute and marginal measures of total impacts complement each other. Absolute values depend on the level of exposure and evaluate adaptation and effects on profits regarding abrupt changes in drivers. Marginal measures show the effect of marginal changes in drivers and are independent of the level of exposure. This is useful when the level of exposure is uncertain. The marginal measures correspond to the slope of the respective absolute measures. Marginals also provide an overview of trade-offs among drivers' effects on profits and adaptations.

Using the dynamics of the bio-economic model, we identify the new steady states for changes in the considered drivers. Table \ref{Tab:steadyStatesParms} shows the new interior steady states for changes in wages ($\omega$), returns to effort ($\epsilon$), and stock harvesting efficiency ($\chi_i$). We find the interior steady states for the upper and lower values presented in Table \ref{Tab:parmsMeaning}. The first row shows the values of stock, quantities, prices, and utility for the initial analysis in steady state.

The lower bound of wages correspond to a decrease of 4\% of the initial value. This is a bifurcation point such that for lower values of $\omega$ there is no interior steady state. Decreases in wages reduce the fishing costs and increase the quantities harvested of plaice; this reduce prices and overall decreases utility. Further reduction in wages causes a collapse of the fishery. The reduced costs of fishing creates incentives to explode the plaice fishery because quotas are not binding for this species. Increases in wages has the highest increase in utility overall drivers in the long term. 

The lower boundary of returns to effort also presents a bifurcation. The decrease in $\epsilon$ presented in Table \ref{Tab:steadyStatesParms} corresponds to 1\% of the initial value, and with lower values, there is no interior steady state. An increase in $\epsilon$ in the long run increases the harvest of plaice and leaves the fishery with only the metier 2. Changes in $\chi_i$ resemble similar dynamics to $\epsilon$. Cases in which $\chi_2$ increases also leave the fishery practicing only the metier 2.


\begin{table}[h!]
\centering
\renewcommand{\arraystretch}{1}
\scalebox{0.7}{
\begin{tabular}{l M{1.4cm}M{1.5cm}M{1.4cm}p{1.4cm}p{1.4cm}p{1.4cm}p{1.4cm}p{1.4cm}}
\hline
\hline
\\[-1.8ex]
& \multicolumn{2}{c}{\textbf{Stocks }} & \multicolumn{2}{c}{\textbf{Quantities }} & \multicolumn{2}{c}{\textbf{Prices }} & \multicolumn{2}{c}{\textbf{Fleet Size}} \\
\\[-1.8ex]
\hline
\\[-1.8ex]
Driver & Plaice & Sole & Plaice& Sole & Plaice  & Sole  & M\'{e}tier1  & M\'{e}tier2  \\
\\[-1.8ex]
\hline
\hline
\\[-1.8ex]
Initial & 586,709 & 85,937 & 132,122 & 17,545 & 3.67 & 6.63 & 674 & 2315\\ 								& Tonnes & Tonnes & Tonnes & Tonnes & Eur/Kg & Eur/Kg &  & \\
\hline
\\[-1.5ex]
Wages-Low & 95\% & 100\% & 100.4\% & 100\% & 99.70\% & 97.98\% & 106\% & 98\% \\
\\[-1.5ex]
Wages-Up & 135\% & 100\% & 91.9\% & 100\% & 106.76\% & 117.05\% & 49\% & 117\% \\
\\[-1.5ex]
$\epsilon_{Lw}$ & 43\% & 100\% & 80.5\% & 100\% & 118.26\% & 91.19\% & 107\% & 92\% \\
\\[-1.5ex]
$\epsilon_{Up}$ & 159\% & 100\% & 83.1\% & 100\% & 114.58\% & 195.56\% & 0\% & 142\% \\
\\[-1.5ex]
$\chi_1$-Lw & 108\% & 100\% & 98.9\% & 100\% & 100.84\% & 71.33\% & 173\% & 71\% \\
\\[-1.5ex]
$\chi_1$-Up & 110\% & 100\% & 98.5\% & 100\% & 101.16\% & 100.00\% & 99\% & 100\% \\
\\[-1.5ex]
$\chi_2$-Lw & 25\% & 100\% & 57.6\% & 100\% & 152.74\% & 86.78\% & 91\% & 87\% \\
\\[-1.5ex]
$\chi_2$-Up & 80\% & 100\% & 99.6\% & 100\% & 98.58\% & 184.11\% & 0\% & 148\% \\
\\[-1.5ex]
\hline
\hline
\end{tabular}
}
\caption[Steady states values of stocks, quantities and prices]{Steady states values of stocks, quantities and prices for changes in the drivers considered. The values are relative to the initial steady state.}
\label{Tab:steadyStatesParms}
\end{table}

\section{Discussion}

%

Some studies assess adaptive capacities of a whole socio-ecological system (SES) \citep{Carpenter2008, Cottrell2020}, and others of a social community embedded in the SES \citep{Chen2020,Cabral2015,Cinner2013}. When assessing adaptive capacities in most cases the unit of analysis is unclear, i.e., ``the adaptive capacity of what to what?'' \citep{Whitney2017}. Our framework answers this question for multiple drivers. We develop a framework that can distinguish various types of adaptation and the degree to which the adaptation counteracts/enhances the impacts of drivers on the system property. The application of this framework to a bio-economic model calibrated to the flatfish German North Sea fishery served as proof of concept to exemplify the use of it. The analysis of the case study demonstrates that the degree of mitigation or enhancement of harmful or beneficial impacts on the system property is driver-dependent. We specifically focus our analysis on fishers' adaptation through effort, illustrating how this adaptation increases profits. We also present the effects of adaptation on quantities and utility within the economy.

Our adaptive effort measure represents the full-time equivalent (FTE) units necessary to perform the fishing activity. It gives an indication of employment changes that this adaptation would cause. With our framework, we distinguish the level of adaptation to different drivers, focusing on returns to effort, wages and stock harvesting efficiency.We identify the magnitude of optimal adaptation by fishers to these drivers and its impact on quantities and utility.. 

In the short term we show that quantities are mostly affected by absolute and marginal changes in returns to effort, followed by stock harvesting efficiency and wages. Regulations aimed at controlling fishing activities have a more substantial impact on quantities compared to regulations directed at changing stock harvesting efficiency.
Monitoring or increasing requirements on reporting the fishing activity have higher decrease in marginal quantities than marginal changes in stock harvesting efficiency. Changes in the latter results from increasing coverage of MPAs \citep{Russi2016}, the current development of off-shore wind farms \citep{Stelzenmuller2020}. These aspects increase the time at sea, making the fishing process less efficient. In absolute terms the effect of fishers adaptation on quantities and utility depends on the level of exposures considered. Larger reductions of the available space to fish can have differentiate effects than changes in fishing monitoring. 

Our analysis show that in the long term policies decreasing the available space to fish (increasing $\chi_i$) could cause the lost of firms targeting plaice (metier 1). Yet, this is the result of our model assuming free entry-exit of firms, however, in reality fishers are constrained by higher costs of new vessels and investments. These costs are strongly influenced by vessel size and age \citep{Lam2011}. The increasing regulations hence could cause a decrease in the number of fishing firms. Further research including restrictions in the fleet size in the model can lead to a better understanding of the system once it is expose to changes in drivers.

The application of our framework also contributes to distinguish effects of drivers on multiple system properties. We evaluate the impacts of changing adaptation to drivers, on quantities and utility in the system. Our findings reveal that adaptation to wages has the lowest effect on utility but a higher level in absolute and marginal quantities. A marginal increase in wages causes a reduction in adaptive effort for almost the same proportion as an increase in stock harvesting efficiency (see Fig \ref{fig:marginalAdaptiveEffort}). This reduction in effort lowers costs, leading to an increase in quantities. These results demonstrate that changes in effort due to a marginal decrease in the space available to fish (increasing $\chi_i$) could be offset by a marginal increase in wages. This perspective provides policymakers with insights into trade-offs among policies, allowing for the consideration of further reductions in effort (Full-Time Equivalent) without generating harmful effects on employment in the sector..  



Our framework operationalizes the concept of  \citet{Ionescu2009}  by eliciting the magnitude of the optimal adaptation that can mitigate or enhance harmful/beneficial impacts in an ecological-economic system. In our case study, we show how for some drivers the optimal adaptation consists in increasing effort and for others to decrease it. This contrasts with the general measures of adaptive capacity that only evaluate adaptation to one driver. \citet{Thiault2019} mention that adaptation strategies aimed to reduce the total impact to one driver may influence impacts on others. 

The design of our framework allows us to distinguish adaptations to changes in drivers that cause harm or benefit to a  system. In this study we exemplify this by evaluating the impacts of changes in drivers from an steady state in positive and negative directions. This distinction is important because the adaptations and impacts on the system  can vary according to the direction of the effect. For instance, in our case study we show that the magnitude of change in profits when wages decrease is higher than the magnitude of change for an increase in wages, i.e, adaptation is higher for a decrease in wages than for an increase. It is relevant to identify these mechanisms because positive impacts caused by a driver can mitigate the harmful impacts of others. \citet{Gallopin2006} also mentions that disturbances in a system can also cause beneficial transformations which need to be addresses in order to have a improved measure of vulnerability.


\section{Limitations}

We develop a framework that allow us to assess and disentangle adaptations, sensitivities and total impacts of multiple drives on a coupled ecological-economic system. We use a bio-economic model to apply this framework because it is the minimum viable product that incorporates the key interconnectedness among economic and ecological sub-systems. In the application of the bio-economic model there are some assumptions inherent to these type of models, however, they are not required to apply the framework. The assumptions of the bio-economic model have an effect on the implications and results of the application.      


Although our model is a oversimplification of the multiple dynamics embedded in this fishery, still provides an understanding of the underlying mechanisms of adaptation affecting quantities and utility. Due to the stylish nature of the model and the complexity of the reality, it limits the results. In this sense this results are also stylized.

\section{Conclusion}

In the multidisciplinary field of adaptation and adaptive capacity, various definitions and concepts exist, contributing to confusion and imprecise policy advice. We have developed a framework that aims to clarify and disentangle sensitivity, total impacts, and adaptation through mathematical modeling. Our framework enables the assessment of adaptations to multiple drivers affecting a system property, facilitating the distinction between the benefits and harms of these drivers on the coupled ecological-economic system.

As a proof of concept we apply our framework to a calibrated bio-economic model of the North Sea flatfish fishery. We investigate the adaptations of fisheries profits to multiple drivers and elicit the optimal adaptation effect in quantities and utility. Among the three drivers evaluated, we identify those that fishers can adapt the best through effort. We find that adaptation to marginal changes in returns to effort generate higher changes in quantities than marginal changes in stock harvesting efficiency. Moreover, changes in effort due to marginal changes in wages have a low impact on utility and a high impact in quantities. The fact that our framework allows us the comparison of adaptation impacts among multiple drivers serves as a departure point to identify trade-offs or counteracting effects among policies.   

Our results exemplify the extent to which various drivers harm or enhance well-being in this fishery and to what extent the fishery can mitigate these effects endogenously. This framework can be applied to other fisheries regions and be used with different bio-economic models. We consider that the generality of the definitions makes the application of our framework easy to implement.


\bigskip

\textbf{Acknowledgments}: We thank Hermman Held for the feedback and useful discussions that enrich this manuscript.

\textbf{Funding Sources}: This work was supported by The German Federal Ministry of Education and Science as part of the SeaUseTip Project [grant number 01LC1825C, 2018].

\newpage
\thispagestyle{plain} 
\mbox{}

\printbibliography

@techreport{STECFsocial2020,
author = {STECF},
booktitle = {Social dimension of the CFP (STECF-20-14)},
institution = {European Commission},
title = {{Social dimension of the CFP (STECF-20-14)}},
url = {https://publications.jrc.ec.europa.eu/repository/handle/JRC123058},
year = {2020}
}

@techreport{EuropeanComission2023,
author = {EC},
booktitle = {Communication From The Commission To The European Parliament, The Council, The European Economic And SocialCommittee And The Committee Of The Regions. EU Action Plan: Protecting and restoring marine ecosystems for sustainable and resilient fisheries},
institution = {European Commission},
title = {{Communication From The Commission To The European Parliament, The Council, The European Economic And SocialCommittee And The Committee Of The Regions. EU Action Plan: Protecting and restoring marine ecosystems for sustainable and resilient fisheries}},
url = {https://eur-lex.europa.eu/legal-content/EN/TXT/?uri=CELEX:52023DC0102},
year = {2023}
}

@article{Nielsen2018,
author = {Nielsen, J. Rasmus and Thunberg, Eric and Holland, Daniel S. and Schmidt, Jorn O. and Fulton, Elizabeth A. and Bastardie, Francois and Punt, Andre E. and Allen, Icarus and Bartelings, Heleen and Bertignac, Michel and Bethke, Eckhard and Bossier, Sieme and Buckworth, Rik and Carpenter, Griffin and Christensen, Asbj{\o}rn and Christensen, Villy and Da-Rocha, Jos{\'{e}} M. and Deng, Roy and Dichmont, Catherine and Doering, Ralf and Esteban, Aniol and Fernandes, Jose A. and Frost, Hans and Garcia, Dorleta and Gasche, Loic and Gascuel, Didier and Gourguet, Sophie and Groeneveld, Rolf A. and Guill{\'{e}}n, Jordi and Guyader, Olivier and Hamon, Katell G. and Hoff, Ayoe and Horbowy, Jan and Hutton, Trevor and Lehuta, Sigrid and Little, L. Richard and Lleonart, Jordi and Macher, Claire and Mackinson, Steven and Mahevas, Stephanie and Marchal, Paul and Mato-Amboage, Rosa and Mapstone, Bruce and Maynou, Francesc and Merz{\'{e}}r{\'{e}}aud, Mathieu and Palacz, Artur and Pascoe, Sean and Paulrud, Anton and Plaganyi, Eva and Prellezo, Raul and van Putten, Elizabeth I. and Quaas, Martin and Ravn-Jonsen, Lars and Sanchez, Sonia and Simons, Sarah and Th{\'{e}}baud, Olivier and Tomczak, Maciej T. and Ulrich, Clara and van Dijk, Diana and Vermard, Youen and Voss, Rudi and Waldo, Staffan},
doi = {10.1111/faf.12232},
issn = {14672979},
journal = {Fish and Fisheries},
number = {1},
pages = {1--29},
title = {{Integrated ecological–economic fisheries models—Evaluation, review and challenges for implementation}},
volume = {19},
year = {2018}
}

@book{Ricker1975,
author = {Ricker, W.E. 1975.},
booktitle = {Bulletin of the Fishers Research Board of Canada},
doi = {10.1038/108070b0},
edition = {Bulletin 1},
issn = {00280836},
number = {2706},
pages = {70},
publisher = {Deaprtment of the Environment Fisheries and Marine Service},
title = {{Computation and interpretation of biological statistics of fish populations}},
volume = {108},
year = {1975}
}

@article{Hufschmidt2011,
author = {Hufschmidt, Gabi},
doi = {10.1007/s11069-011-9823-7},
issn = {0921030X},
journal = {Natural Hazards},
number = {2},
pages = {621--643},
title = {{A comparative analysis of several vulnerability concepts}},
volume = {58},
year = {2011}
}

@article{Cinner2013,
author = {Cinner, Joshua E. and Huchery, Cindy and Darling, Emily S. and Humphries, Austin T. and Graham, Nicholas A.J. and Hicks, Christina C. and Marshall, Nadine and McClanahan, Tim R.},
doi = {10.1371/journal.pone.0074321},
issn = {19326203},
journal = {PloS one},
number = {9},
title = {{Evaluating social and ecological vulnerability of coral reef fisheries to climate change.}},
volume = {8},
year = {2013}
}

@article{ICES2019,
author = {ICES},
doi = {10.17895/ices.pub.4613},
journal = {ICES Advice on fishing opportunities, catch, and effort Greater North Sea ecoregion},
number = {June 2019},
pages = {1--11},
title = {{Plaice (Pleuronectes platessa) in Subarea 4 (North Sea) and Subdivision 20 (Skagerrak) ICES advice on fishing opportunities}},
volume = {4},
year = {2019}
}

@article{Adger2006,
author = {Adger, W. Neil},
doi = {10.1016/j.gloenvcha.2006.02.006},
issn = {09593780},
journal = {Global Environmental Change},
number = {3},
pages = {268--281},
title = {{Vulnerability}},
volume = {16},
year = {2006}
}

@book{Quante2016,
author = {Quante, Markus and Colijn, Franciscus},
doi = {10.1007/978-3-319-39745-0},
isbn = {9783319397436},
number = {October},
pages = {11--14},
title = {{North Sea Region Climate Assessment}},
year = {2016}
}

@techreport{IPCC2001,
author = {IPCC},
booktitle = {Change},
institution = {Intergovernmental Panel on Climate Change},
title = {{T ECHNICAL S UMMARY C LIMATE C HANGE 2001 : M ITIGATION A Report of Working Group III of the Intergovernmental Panel on Climate Change}},
url = {http://www.grida.no/climate/ipcc{\_}tar/wg2/ pdf/wg2TARtechsum.pdf},
year = {2001}
}

@article{Cabral2015,
author = {Cabral, P. and Levrel, H. and Schoenn, J. and Thi{\'{e}}baut, E. and {Le Mao}, P. and Mongruel, R. and Rollet, C. and Dedieu, K. and Carrier, S. and Morisseau, F. and Daures, F.},
doi = {10.1016/j.ecoser.2014.09.007},
issn = {22120416},
journal = {Ecosystem Services},
pages = {306--318},
publisher = {Elsevier},
title = {{Marine habitats ecosystem service potential: A vulnerability approach in the Normand-Breton (Saint Malo) Gulf, France}},
url = {http://dx.doi.org/10.1016/j.ecoser.2014.09.007},
volume = {16},
year = {2015}
}

@article{ICES2015,
author = {ICES},
journal = {ICES Advice on fishing opportunities, catch, and effort},
number = {November 2019},
pages = {1--9},
title = {{Sole (Solea solea) in Subarea IV (North Sea)}},
volume = {4},
year = {2019}
}

@article{Daan1997,
author = {Daan, Niels},
doi = {10.1016/S1385-1101(97)00026-9},
issn = {13851101},
journal = {Journal of Sea Research},
number = {3-4},
pages = {321--341},
title = {{TAC management in North Sea flatfish fisheries}},
volume = {37},
year = {1997}
}

@incollection{Pinnegar2016,
author = {Pinnegar, John K. and Engelhard, Georg H. and Jones, Miranda C. and Cheung, William W.L. and Peck, Myron A. and Rijnsdorp, Adriaan D and Brander, Keith},
booktitle = {North Sea Region Climate Change Assessment},
chapter = {12},
doi = {10.1007/978-3-319-39745-0},
editor = {Quante, Markus and Colijn, Franciscus},
isbn = {ISBN 978-3-319-39743-6},
pages = {375--396},
publisher = {Springer Open},
title = {{Socio-economic Impacts—Fisheries}},
year = {2016}
}

@article{Berrouet2018,
author = {Berrouet, Lina Mar{\'{i}}a and Machado, Jenny and Villegas-Palacio, Clara},
doi = {10.1016/j.ecolind.2017.07.051},
issn = {1470160X},
journal = {Ecological Indicators},
number = {September 2017},
pages = {632--647},
title = {{Vulnerability of socio—ecological systems: A conceptual Framework}},
volume = {84},
year = {2018}
}

@book{Bene2012,
author = {B{\'{e}}n{\'{e}}, Christophe and Wood, Rachel Godfrey and Newsham, Andrew and Davies, Mark},
booktitle = {IDS Working Papers},
doi = {10.1111/j.2040-0209.2012.00405.x},
isbn = {9781781180914},
number = {405},
pages = {1--61},
title = {{Resilience: New Utopia or New Tyranny? Reflection about the Potentials and Limits of the Concept of Resilience in Relation to Vulnerability Reduction Programmes}},
volume = {2012},
year = {2012}
}

@article{Stelzenmuller2020,
author = {Stelzenm{\"{u}}ller, Vanessa and Gimpel, Antje and Letschert, Jonas and Kraan, Casper and D{\"{o}}ring, Ralf},
journal = {European Parliament, Policy Department for Structural and Cohesion Policies, Brussels. Research for PECH Committee - European Parliament, Policy Department for Structural and Cohesion policies},
number = {October},
pages = {104},
title = {{Impact of the use of offshore wind and other marine renewables on European fisheries}},
url = {https://www.europarl.europa.eu/thinktank/en/document.html?reference=IPOL{\_}STU(2020)652212},
year = {2020}
}

@article{Lam2011,
author = {Lam, Vicky W.Y. and Sumaila, Ussif Rashid and Dyck, Andrew and Pauly, Daniel and Watson, Reg},
doi = {10.1093/icesjms/fsr121},
issn = {10543139},
journal = {ICES Journal of Marine Science},
number = {9},
pages = {1996--2004},
title = {{Construction and first applications of a global cost of fishing database}},
volume = {68},
year = {2011}
}

@article{Etherton2015,
author = {Etherton, Mark},
doi = {10.1016/j.fishres.2015.05.009},
issn = {01657836},
journal = {Fisheries Research},
pages = {76--81},
publisher = {Elsevier B.V.},
title = {{European plaice (Pleuronectes platessa) and sole (Solea solea) indirect age validation using otoliths from mark-recapture experiments from the North Sea}},
url = {http://dx.doi.org/10.1016/j.fishres.2015.05.009},
volume = {170},
year = {2015}
}

@article{McDowell2012,
author = {McDowell, Julia Z. and Hess, Jeremy J.},
doi = {10.1016/j.gloenvcha.2011.11.002},
issn = {09593780},
journal = {Global Environmental Change},
number = {2},
pages = {342--352},
publisher = {Elsevier Ltd},
title = {{Accessing adaptation: Multiple stressors on livelihoods in the Bolivian highlands under a changing climate}},
url = {http://dx.doi.org/10.1016/j.gloenvcha.2011.11.002},
volume = {22},
year = {2012}
}

@techreport{stecf2019,
author = {STECF, European Commission},
doi = {10.2788/72742},
isbn = {9789279152115},
pages = {7--11},
title = {{Scientific , Technical and Economic Committee for Fisheries ( STECF ) Report of the Working Group on the review of national reports on Member States efforts to achieve balance between fleet capacity and fishing opportunities Subgroup on Balance between re}},
url = {https://stecf.jrc.ec.europa.eu/dd/fleet},
year = {2019}
}

@book{Clark1990,
author = {Clark, Colin W},
booktitle = {John Wiley {\&} Sons},
number = {1},
publisher = {John Wiley {\&} Sons},
title = {{Mathematical Bioeconomics: The Optimal Management of Renewable Resources}},
volume = {60},
year = {1990}
}

@book{STECF2021,
author = {STECF},
doi = {10.2760/60996},
editor = {Prellezo, Ra{\'{u}}l and Carvallo, Natacha and Virtanen, Jarno and Guillen, Jordi},
isbn = {978-92-76-40959-5},
number = {July},
pages = {302p},
title = {{Scientific, technical and economic committee for fisheries (STECF) – The 2013 annual economic report on the EU fishing fleet (STECF-13-15)}},
url = {http://stecf.jrc.ec.europa.eu/documents/43805/581354/2013-09{\_}STECF+13-15+-+AER+EU+Fleet+2013{\_}JRC84745.pdf},
year = {2021}
}

@techreport{IPCC2022,
author = {IPCC},
doi = {10.1017/9781009325844.Front},
institution = {Intergovernmental Panel on Climate Change},
isbn = {9781009325844},
title = {{Climate Change 2022 : Impacts , Adaptation and Vulnerability Working Group II Contribution to the Sixth Assessment Report of the Intergovernmental Panel on Climate Change. [H.-O. P{\"{o}}rtner, D.C. Roberts, M. Tignor, E.S. Poloczanska, K. Mintenbeck, A. Alegr{\'{i}}}},
url = {https://report.ipcc.ch/ar6/wg2/IPCC{\_}AR6{\_}WGII{\_}FullReport.pdf},
year = {2022}
}

@techreport{EUMOFA2013,
author = {EUMOFA},
booktitle = {European Commision},
institution = {European Commision},
number = {January},
title = {{Guidelines Data and methodology for price structure analysis}},
year = {2013}
}

@article{Thiault2019,
author = {Thiault, Lauric and Gelcich, Stefan and Cinner, Joshua E. and Tapia‐Lewin, Sebastian and Chlous, Fr{\'{e}}d{\'{e}}rique and Claudet, Joachim},
doi = {10.1002/pan3.10056},
issn = {2575-8314},
journal = {People and Nature},
number = {4},
pages = {573--589},
title = {{Generic and specific facets of vulnerability for analysing trade‐offs and synergies in natural resource management}},
volume = {1},
year = {2019}
}

@article{Luers2003,
author = {Luers, Amy L and Lobell, David B and Sklar, Leonard S and Addams, C. Lee and Matson, Pamela A},
doi = {10.1016/S0959-3780(03)00054-2},
isbn = {6507251992},
issn = {09593780},
journal = {Global Environmental Change},
number = {4},
pages = {255--267},
title = {{A method for quantifying vulnerability, applied to the agricultural system of the Yaqui Valley, Mexico}},
volume = {13},
year = {2003}
}

@techreport{ICES2021b,
author = {ICES},
booktitle = {ICES Advice on fishing opportunities, catch, and effort},
number = {May},
pages = {1--9},
title = {{Sole ( Solea solea ) in Subarea 4 ( North Sea )}},
volume = {4},
year = {2021}
}

@article{Russi2016,
author = {Russi, Daniela and Pantzar, Mia and Kettunen, Marianne and Gitti, Giulia and Mutafoglu, Konar and {Patrick Ten Brink}, Monika Kotulak},
journal = {Institute for European Environmental Policy (IEEP)},
number = {May},
pages = {1--92},
title = {{Socio-economic benefits of the EU marine protected areas}},
url = {https://ec.europa.eu/environment/nature/natura2000/marine/docs/Socio -Economic Benefits of EU MPAs.pdf},
year = {2016}
}

@article{Aarts2009,
author = {Aarts, G. and Poos, J. J.},
doi = {10.1093/icesjms/fsp033},
issn = {10543139},
journal = {ICES Journal of Marine Science},
number = {4},
pages = {763--771},
title = {{Comprehensive discard reconstruction and abundance estimation using flexible selectivity functions}},
volume = {66},
year = {2009}
}

@article{Whitney2017,
author = {Whitney, Charlotte K. and Bennett, Nathan J. and Ban, Natalie C. and Allison, Edward H. and Armitage, Derek and Blythe, Jessica L. and Burt, Jenn M. and Cheung, William and Finkbeiner, Elena M. and Kaplan-Hallam, Maery and Perry, Ian and Turner, Nancy J. and Yumagulova, Lilia},
doi = {10.5751/ES-09325-220222},
issn = {17083087},
journal = {Ecology and Society},
number = {2},
title = {{Adaptive capacity: From assessment to action in coastal social-ecological systems}},
volume = {22},
year = {2017}
}

@techreport{ICES2021a,
author = {ICES},
booktitle = {ICES Advice on fishing opportunities, catch, and effort},
institution = {ICES},
number = {ICES Advice 2019},
pages = {1--11},
title = {{Plaice (Pleuronectes platessa) in Subarea 4 (North Sea) and Subdivision 20 (Skagerrak)}},
url = {https://doi.org/10.17895/ices.advice.5644},
volume = {4},
year = {2021}
}

@article{EuropeanCommission2014,
author = {{European Commission}},
journal = {Official journal of the European Union},
number = {Oktober},
pages = {35--39},
title = {{COMMISSION DELEGATED REGULATION (EU) No 1395/2014 of 20 October 2014 - establishing a discard plan for certain small pelagic fisheries and fisheries for industrial purposes in the North Sea}},
volume = {2014},
year = {2014}
}

@misc{sole,
author = {Wikipedia},
title = {{Lined sole}},
url = {https://commons.wikimedia.org/wiki/File:Lined{\_}sole.jpg},
urldate = {2021-01-26}
}

@article{Serfilippi2018,
author = {Serfilippi, Elena and Ramnath, Gayatri},
doi = {10.1111/apce.12202},
issn = {14678292},
journal = {Annals of Public and Cooperative Economics},
number = {4},
pages = {645--664},
title = {{Resilience Measurement and Conceptual Frameworks: a Review of the Literature}},
volume = {89},
year = {2018}
}

@article{Nash2019,
author = {Nash, John C. and Varadhan, Ravi and Grothendieck, Gabor},
journal = {CRAN},
title = {{Package ‘ optimr ': A Replacement and Extension of the 'optim' Function}},
year = {2019}
}

@article{Dixit1977,
author = {Dixit, Avinash K. and Stiglitz, Joseph E.},
journal = {The American Economic Review},
number = {3},
pages = {297--308},
title = {{Monopolistic Competition and Optimum Product Diversity}},
volume = {67},
year = {1977}
}

@article{Engelhard2011,
author = {Engelhard, Georg H. and Pinnegar, John K. and Kell, Laurence T. and Rijnsdorp, Adriaan D.},
doi = {10.1093/icesjms/fsr031},
issn = {10543139},
journal = {ICES Journal of Marine Science},
number = {6},
pages = {1090--1104},
title = {{Nine decades of North Sea sole and plaice distribution}},
volume = {68},
year = {2011}
}

@article{Carpenter2008,
author = {Carpenter, Stephen R. and Brock, William A.},
doi = {10.5751/ES-02716-130240},
issn = {17083087},
journal = {Ecology and Society},
number = {2},
title = {{Adaptive capacity and traps}},
volume = {13},
year = {2008}
}

@book{IPCC2014a,
author = {IPCC},
doi = {https://doi.org/10.1017/CBO9781107415379},
isbn = {9781107641655},
pages = {1132},
publisher = {Cambridge University Press},
title = {{Climate Change 2014 – Impacts, Adaptation and Vulnerability: Part A: Global and Sectoral Aspects: Working Group II Contribution to the IPCC Fifth Assessment Report}},
year = {2014}
}

@article{Grafton2019,
author = {Grafton, R. Quentin and Doyen, Luc and B{\'{e}}n{\'{e}}, Christophe and Borgomeo, Edoardo and Brooks, Kate and Chu, Long and Cumming, Graeme S. and Dixon, John and Dovers, Stephen and Garrick, Dustin and Helfgott, Ariella and Jiang, Qiang and Katic, Pamela and Kompas, Tom and Little, L. Richard and Matthews, Nathanial and Ringler, Claudia and Squires, Dale and Steinshamn, Stein Ivar and Villasante, Sebasti{\'{a}}n and Wheeler, Sarah and Williams, John and Wyrwoll, Paul R.},
doi = {10.1038/s41893-019-0376-1},
issn = {23989629},
journal = {Nature Sustainability},
number = {10},
pages = {907--913},
publisher = {Springer US},
title = {{Realizing resilience for decision-making}},
url = {http://dx.doi.org/10.1038/s41893-019-0376-1},
volume = {2},
year = {2019}
}

@article{Knijn1993,
author = {Knijn, Ruud J and Boon, Trevor W and Heessen, Henk J L and Hislop, John R G},
number = {194},
pages = {268 p.},
title = {{Atlas of the North Sea fishes}},
year = {1993}
}

@article{Reed2013,
author = {Reed, M. S. and Podesta, G. and Fazey, I. and Geeson, N. and Hessel, R. and Hubacek, K. and Letson, D. and Nainggolan, D. and Prell, C. and Rickenbach, M. G. and Ritsema, C. and Schwilch, G. and Stringer, L. C. and Thomas, A. D.},
doi = {10.1016/j.ecolecon.2013.07.007},
issn = {09218009},
journal = {Ecological Economics},
pages = {66--77},
publisher = {The Authors},
title = {{Combining analytical frameworks to assess livelihood vulnerability to climate change and analyse adaptation options}},
url = {http://dx.doi.org/10.1016/j.ecolecon.2013.07.007},
volume = {94},
year = {2013}
}

@article{Thiault2019a,
author = {Thiault, Lauric and Mora, Camilo and Cinner, Joshua E. and Cheung, William W.L. and Graham, Nicholas A.J. and Januchowski-Hartley, Fraser A. and Mouillot, David and {Rashid Sumaila}, U. and Claudet, Joachim},
doi = {10.1126/sciadv.aaw9976},
issn = {23752548},
journal = {Science Advances},
number = {11},
pages = {1--10},
pmid = {31807697},
title = {{Escaping the perfect storm of simultaneous climate change impacts on agriculture and marine fisheries}},
volume = {5},
year = {2019}
}

@article{Ionescu2009,
author = {Ionescu, Cezar and Klein, Richard J.T. and Hinkel, Jochen and {Kavi Kumar}, K. S. and Klein, Rupert},
doi = {10.1007/s10666-008-9179-x},
issn = {14202026},
journal = {Environmental Modeling and Assessment},
number = {1},
pages = {1--16},
title = {{Towards a formal framework of vulnerability to climate change}},
volume = {14},
year = {2009}
}

@misc{plaice,
author = {Wikipedia},
title = {{Hippoglossoides platessoidestle}},
url = {https://commons.wikimedia.org/wiki/File:Hippoglossoides{\_}platessoides.jpg},
urldate = {2021-01-26}
}

@article{Gattuso2018,
author = {Gattuso, Jean Pierre and Magnan, Alexandre K. and Bopp, Laurent and Cheung, William W.L. and Duarte, Carlos M. and Hinkel, Jochen and Mcleod, Elizabeth and Micheli, Fiorenza and Oschlies, Andreas and Williamson, Phillip and Bill{\'{e}}, Rapha{\"{e}}l and Chalastani, Vasiliki I. and Gates, Ruth D. and Irisson, Jean Olivier and Middelburg, Jack J. and P{\"{o}}rtner, Hans Otto and Rau, Greg H.},
doi = {10.3389/fmars.2018.00337},
issn = {22967745},
journal = {Frontiers in Marine Science},
number = {OCT},
title = {{Ocean solutions to address climate change and its effects on marine ecosystems}},
volume = {5},
year = {2018}
}

@article{VanKeeken2007,
author = {van Keeken, O. A. and van Hoppe, M. and Grift, R. E. and Rijnsdorp, A. D.},
doi = {10.1016/j.seares.2006.09.002},
issn = {13851101},
journal = {Journal of Sea Research},
number = {2-3 SPEC. ISS.},
pages = {187--197},
title = {{Changes in the spatial distribution of North Sea plaice (Pleuronectes platessa) and implications for fisheries management}},
volume = {57},
year = {2007}
}

@article{Quaas2013b,
author = {Quaas, Martin F and Requate, Till},
doi = {10.1111/sjoe.12002},
journal = {The Sacandinavian Journal of Economics},
number = {2},
pages = {381--422},
title = {{Sushi or Fish Fingers ? Seafood Diversity , Collapsing Fish Stocks , and Multispecies Fishery Management}},
volume = {115},
year = {2013}
}

@article{Chen2020,
author = {Chen, Qi and Su, Hongyan and Yu, Xuan and Hu, Qiuguang},
doi = {10.3390/ijerph17030765},
issn = {16604601},
journal = {International Journal of Environmental Research and Public Health},
number = {3},
pages = {1--17},
title = {{Livelihood vulnerability of marine fishermen to multi-stresses under the vessel buyback and fishermen transfer programs in China: The case of Zhoushan City, Zhejiang Province}},
volume = {17},
year = {2020}
}

@article{Folke2016,
author = {Folke, Carl},
doi = {https://doi.org/10.5751/ES-09088-210444 Invited},
isbn = {9780199389414},
journal = {Ecology and Society},
number = {4},
pages = {21(4):44},
title = {{Resilience ( Republished )}},
volume = {21},
year = {2016}
}

@article{Prellezo2012,
author = {Prellezo, Ra{\'{u}}l and Accadia, Paolo and Andersen, Jesper L. and Andersen, Bo S. and Buisman, Erik and Little, Alyson and Nielsen, J. Rasmus and Poos, Jan Jaap and Powell, Jeff and R{\"{o}}ckmann, Christine},
doi = {10.1016/j.marpol.2011.08.003},
issn = {0308597X},
journal = {Marine Policy},
number = {2},
pages = {423--431},
title = {{A review of EU bio-economic models for fisheries: The value of a diversity of models}},
volume = {36},
year = {2012}
}

@article{Schuhbauer2016,
author = {Schuhbauer, Anna and Sumaila, U. Rashid},
doi = {10.1016/j.ecolecon.2016.01.018},
issn = {09218009},
journal = {Ecological Economics},
pages = {69--75},
title = {{Economic viability and small-scale fisheries - A review}},
volume = {124},
year = {2016}
}

@article{Cottrell2020,
author = {Cottrell, Stuart and Mattor, Katherine M. and Morris, Jesse L. and Fettig, Christopher J. and McGrady, Pavlina and Maguire, Dorothy and James, Patrick M.A. and Clear, Jennifer and Wurtzebach, Zach and Wei, Yu and Brunelle, Andrea and Western, Jessica and Maxwell, Reed and Rotar, Marissa and Gallagher, Lisa and Roberts, Ryan},
doi = {10.1007/s11625-019-00736-2},
isbn = {0123456789},
issn = {18624057},
journal = {Sustainability Science},
number = {2},
pages = {555--567},
publisher = {Springer Japan},
title = {{Adaptive capacity in social–ecological systems: a framework for addressing bark beetle disturbances in natural resource management}},
volume = {15},
year = {2020}
}

@article{Gallopin2006,
author = {Gallop{\'{i}}n, Gilberto C.},
doi = {10.1016/j.gloenvcha.2006.02.004},
issn = {09593780},
journal = {Global Environmental Change},
number = {3},
pages = {293--303},
title = {{Linkages between vulnerability, resilience, and adaptive capacity}},
volume = {16},
year = {2006}
}

@article{Blanz2019,
author = {Blanz, Benjamin},
doi = {10.1007/s10750-018-3799-1},
isbn = {0123456789},
issn = {1573-5117},
journal = {Hydrobiologia},
number = {845},
pages = {129--144},
publisher = {Springer International Publishing},
title = {{Modelling interactions of fish , fishers and consumers : should bycatch be taken into account ?}},
url = {https://doi.org/10.1007/s10750-018-3799-1},
volume = {1},
year = {2019}
}

\newpage  
\appendix 
  
\section{APPENDIX}
\label{sec:appendix1}
\textbf{Model Description}
\renewcommand{\theequation}{A.\arabic{equation}}
\setcounter{equation}{0}
\renewcommand\thefigure{A.\arabic{figure}} 
\setcounter{figure}{0}

We present a bio-economic model based on \citet{Blanz2019}. It provides us with tools to understand the North Sea fishery complexity. We add to this bio-economic model two main components. First, a variable that accounts for the weight of each species in the household' utility function ($\beta_i$). Second, the logistic growth function was replaced by the Ricker-recruitment function that, to our knowledge and data, provides a better fit to the stock growth for plaice and sole in the North Sea.

\begin{figure}[h]
\begin{center}
  \includegraphics[width=0.9\textwidth]{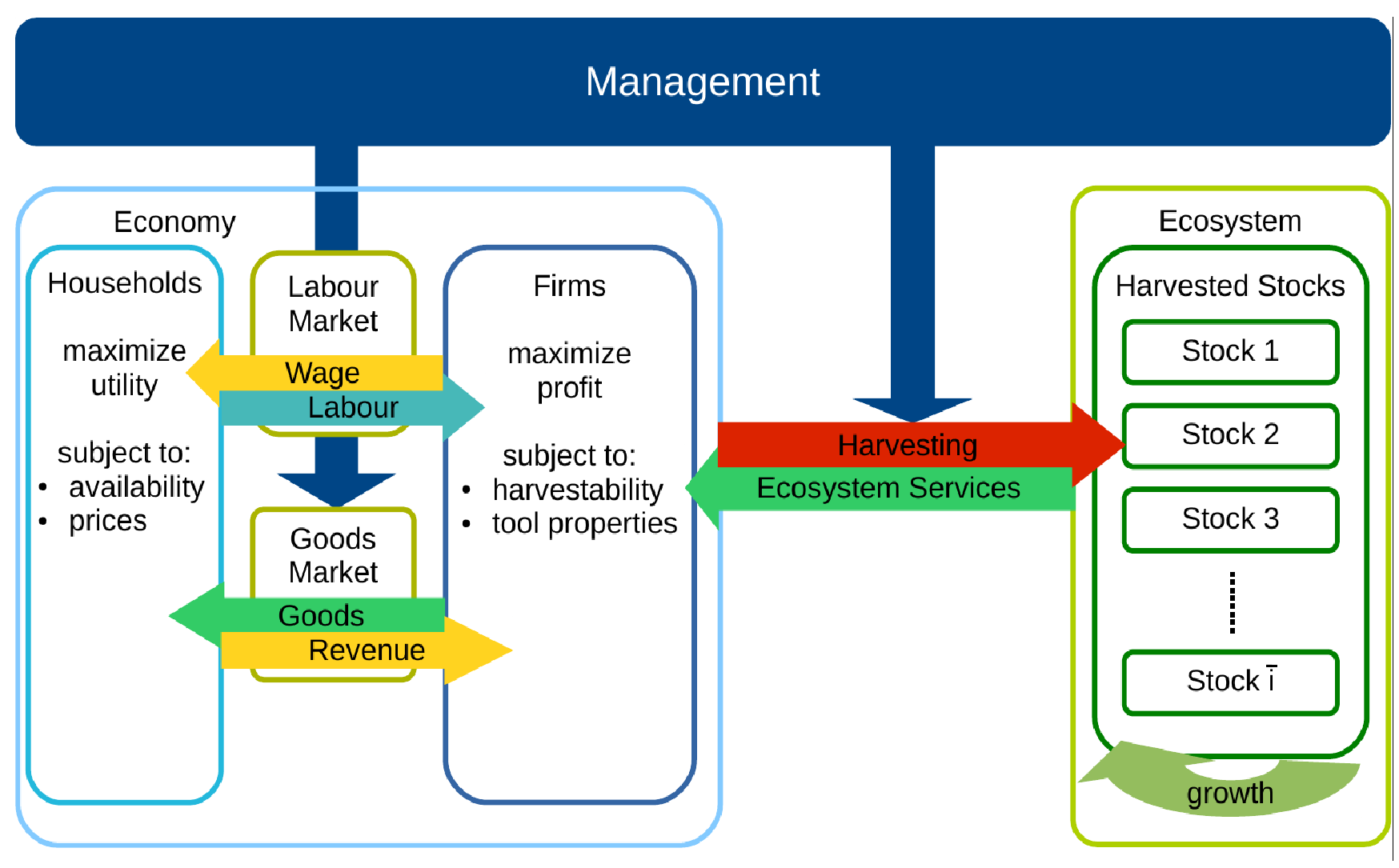}
  \caption {Components of the bio-economic model and their interactions}
  \label{fig:flowchart}
  \end{center}
\end{figure}

Figure \ref{fig:flowchart} shows the components of the model.  An ecosystem component describing the current state and dynamics, harvesting firms maximizing profits, and consumers maximizing contemporaneous utility. The market between the harvesting firms and households allows to sale harvested ecosystem stocks to consumers. The prices on this market and corresponding harvested quantities are determined endogenously. A second labor market allows firms to employ the labor provided by households in the harvesting or manufacturing of a numeraire commodity. Hence, it provides income to households to pay for the fish and other products consumed.

\newpage

\textbf{Ecosystem Properties}

This sub-system is composed of $\bar{i}$ species. Stocks are denoted by $x$ with indexes for species $i\in{I} $, where $I$ is the set of all species $I = [1,i]  \cap {Z}$. Species are assumed to grow each period $t$ due to intrinsic growth $g_{it}$ and are diminished by harvests $H_{it}$. This change in stocks is modeled by differential equations, determining the dynamics of the model. This is the only component of the model that account for time dependency.

 \begin{equation}
  \dot{x_{it}} = g_{it} (\textbf{x}_{t}) - H_{it}  
  \label{eqn:stockChange}
 \end{equation}

In equation \ref{eqn:stockChange} $g_{it}$ is the biomass growth function represented by the Ricker-recruitment growth \ref{eqn:rickerfunction}. 
It depends on the entire vector of stocks, and the parameters $a_i$ and $b_i$. ‘$a_i$' is density independent parameter  proportional to fecundity and ‘$b_i$' is a density-dependent parameter. If density-dependence in the stock-recruitment (growth) relationship does not exist, then $b = 0$. 
 
 \begin{equation}
  g_{i}(\textbf{x}) = a_i(x_i)e^{-b_i x_i}
  \label{eqn:rickerfunction} 
 \end{equation}

\textbf{Harvesting Properties}:

Once the stock for each period is assessed, fisheries make their harvest choices based on the stock available $x_i$. The harvest component includes $\bar{k}$ m\`{e}tiers, which encompasses all that is necessary for the fisher to harvest and is not dependent on the effort i.e. all upfront investments that are necessary to start operating.

M\'etiers are indexed by $k \in{K}$, where $K$ is the set of all m\`{e}tiers $K = [1, k] \cap{Z}$. Each m\'etier has a target species, but may also catch other species, as by-catch. While individual firms may not change their m\'etier, the economy-wide fleet size for each m\'etier is dynamic. The change of gear in use occurs through the market entry and exit of firms performing different m\'etier. where $\bar{i}=\bar{k}$.

Total harvest in the economy $H_{i}$ of species $i$ is determined by the number of firms $n_k$ practicing m\'etier $k$ and the sum of the harvested quantity by each firm $h_{ik}$  targeting the species $i$ with  m\'etier $k$. 

 \begin{equation}
  H_{i} = \sum\limits_{k=1}^{\bar{k}}  n_{k}h_{ik}(e_{k}, x_{i}) 
 \end{equation}

The harvest per firm is defined following the generalized Gordon–Schaefer production function \citep{Clark1990}. Using the m\'etier $k$ the fisher can target the species $i$, but can also harvest other species. The fisher can not control the fish species that she catches. Therefore, the total amount of harvest $H_{i}$ depends  on the effort $e_{k}$ practicing all the m\`{e}tiers $k$ capable of catching that species  ($k \in {K| \nu_{ik} >0} $). The effort experiences diminishing returns to effort $\epsilon$ and is determined under the assumption of perfect markets for harvesting goods and labor. The gear effect is governed by the gear matrix $\nu_{ik}$. The elements of $\nu_{ik}$ specify the catchability for each species $i$ by m\`{e}tier $k$. Species abundance influences the harvest returns per effort through the harvestability function $\chi_{i}(x_{i})$. It captures changes in harvest yield due to changing stocks. Less abundant species are more difficult to catch compared to species with high stock levels $\chi_{i}(x_{i})= x_{i} ^ {\chi_{i}} $. In the following $\chi_{i}(x_{i})$ will be abbreviated as $\chi_{i}$. It specifies a square matrix containing the $\chi_{i}$ along the diagonal and zeros off the diagonal.

 \begin{equation}
  h_{ik}(e_{k}, x_{i}) = \nu_{ik} e_{k}^\epsilon \chi(x_{i}) 
    \label{eqn:harvestspermetierAppen}  
 \end{equation}

The profits of each firm are defined as the difference between income and costs. The income is derived from the quantity of fish harvested $h_{ik}$ times the price of the species $i$, $p_{i}$. Costs include wages $\omega$ times the effort $e_{k}$, which is measured in units of labor, keeping the structure given by \citet{Quaas2013b}. Fixed costs $\phi_{k}$ are defined per m\`{e}tier $k$ and represent fees for entering the markets, fixed price for quotas or also initial capital. In order to maximize profits each firm takes stock levels $x_{i}$, prices $p_{i}$ and wages $\omega$ as given to define their effort $e_{k}$.

 \begin{equation}
  \max_{e_k} \:   \pi_{k} = \sum\limits_{i=1}^{\bar{i}}  h_{ik}(e_{k}, x_{i})p_{i} - \omega e_{k} - \phi_{k}  
  \label{eqn:maxFirms}
  \end{equation}

The maximization of these profits and the assumption of perfect markets leads the firms' profits to zero. Under an open-access scenario it derives to the optimal effort level, given by \eqref{eqn:optEffortZPC}.
Then, the firms' m\`{e}tier specific equilibrium harvest is obtained from replacing $e_{k}^{*}$ in the harvesting production function:

 \begin{equation}
  h_{ik}(x_{i}) = \nu_{ik} e_{k}^ {*\epsilon} \chi(x_{i}) 
 \end{equation}

\textbf{Household Properties}:

The household preferences involve the fish' consumers who have preferences for fish $Q$, and a numeraire commodity $y$. The utility is described by the function:

\begin{equation}
  U(Q,y)=\begin{cases}
    y + \alpha \frac{\eta}{\eta - 1} Q^{\frac{\eta -1}{\eta}}  & \text{for $\eta \neq 1$}.\\
    y + \alpha \ln Q  & \text{for $\eta = 1$}.
  \end{cases}
\end{equation}

The parameter $\eta$ indicates the constant demand elasticity of fish, $\alpha \geq 0$ characterize the importance of fish consumption in overall consumption. Regarding the preferences over the fish species, they are modeled using a Dixit-Stiglitz utility function \citep{Dixit1977}.

\begin{equation}
  Q = Q(\textbf{q}) = \left(  \sum\limits_{i=1}^{\bar{i}} (\beta_{i} q_{i})^{\frac{\sigma -1 }{\sigma}} \right)^{\frac{\sigma}{\sigma - 1}}
  \label{eqn:quantityDemanded}
\end{equation}

In equation \ref{eqn:quantityDemanded}, $q_i$ corresponds to the quantity of the fish species $i$ consumed by the household. $\beta_{i}$ represents the weight of each species in the utility function. This allows us to account for differences in demand quantity for a specific type of fish species. $\sigma > 0$ measures the elasticity of substituting between consumption levels of different species. Hence, perfect substitution is achieved when $\sigma$ tends to infinity ($\sigma \to \infty $), and lower values illustrate the limited substitutability of fish species in consumption. 

The households maximize their utility subject to the budget constrain. They allocate their wages $\omega$ received from providing labor to the fisheries and manufactured sector. The first part of $\omega$ is spent in a manufactured good $y$, which price is normalized to one. A second part is spent in fish, with the amount consumed $q_i$ given the weight of each species in the utility function $\beta_i$ and the price per species $p_i$.

\begin{equation}
 \omega = y + \sum \limits_{i=1}^{\bar{i}} (\beta_i q_i)p_{i} 
 \label{eqn:wages}  
\end{equation}

To keep the analysis tractable, no savings or other capital accumulation is possible in the model. Additionally, Further to what is presented in \citet{Quaas2013b} and following \citet{Blanz2019}, household demand presents an additional restriction called the market-clearing condition. It states that whatever is harvested will be consumed for each species, such that the number of firms are non-negative $n_k \geq 0$

\begin{equation}
 q_i = H_i =\sum \limits_{k=1}^{\bar{k}} n_k h_{ik}(x_{i})
 \label{eqn:marketclearing}  
\end{equation}

\textbf{Firm Optimization Problem}

The firms maximize their profit and therefore find their optimal effort, resulting in the first order condition, from \eqref{eqn:maxFirms}:

\begin{equation}
  \frac{\delta \pi_{k}}{\delta e_{k}} = \epsilon  \left( \sum\limits_{i=1}^{\bar{i}} \nu_{ik} \chi(x_{i}) p_{i} \right) e_{k}^ {\epsilon -1 } - \omega  = 0 
\end{equation}

\begin{equation}
e_{k}^{**} = \left(   \frac{\epsilon}{\omega} \sum\limits_{i=1}^{\bar{i}} \nu_{ik} x_{i}^{\chi_i} p_{i} ) \right) ^\frac{1}{1 - \epsilon} 
  \label{eqn:optEffortBefore}
\end{equation}

Given the assumption of perfect markets in the model, the market pressure on each firm drives profits to zero, what leads into the zero profit condition $\pi_{k} = 0$. Replacing \eqref{eqn:optEffortBefore} in the zero profit condition, we have:

\begin{equation}
 e_{k}^{*} = \frac{\phi_k}{\omega} \frac{\epsilon}{(1 - \epsilon)}
  \label{eqn:optEffortZPC}
\end{equation}
 
This zero profit condition also allows to derive the prices. For this purpose the assumption of $\bar{i}=2$ and $\bar{k}=2$ holds, so that a theoretical solution can be determined. The specific step by step can be found in the appendix of \citet{Blanz2019}.

 Hence,  we have:
 
\begin{equation}
 {}^{p}b_k = \phi_k \left(1 + \frac{\epsilon}{1- \epsilon} \right) \left(\frac{\phi_k}{\omega} \frac{\epsilon}{1 - \epsilon} \right)^{-\epsilon}
\end{equation}
 
\begin{equation}
p_1^* = (\chi_1)^{-1} (\nu_{11} \nu_{22} - \nu_{12} \nu_{21})^{-1} (\nu_{22}{}^{p}b_1  - \nu_{21} {}^{p}b_2)
\end{equation}
 
\begin{equation}
p_2^* = (\chi_2)^{-1} (\nu_{11} \nu_{22} - \nu_{12} \nu_{21})^{-1} (\nu_{11}{}^{p}b_2  - \nu_{12} {}^{p}b_1)
\end{equation}

\textbf{Household Optimization Problem}

The households maximize their utility and choose their quantities $Q$, and $y$.

 \begin{equation}
 \max_{Q, y} \: U (Q,y) \: s.t. \:  \omega = y + \sum \limits_{i=1}^{\bar{i}} (\beta_i q_i)p_{i} 
 \label{eqn:utility} 
  \end{equation}

Solving this maximization problem, lead us to the quantities $q_i^{*}$ demanded by consumers, and $p_i^{*}$ willingness to pay for the fish. This function relates the amount of each species demanded (and consumed) to the prices of all available species. 

\begin{equation}
 q_i^{*} = \alpha^\eta p_i ^{-\sigma} \beta_i^{\sigma-1} \left(\sum_{i'}^{\bar{i}} (p_{i}\beta_i)^{1 - \sigma} \right)^{\frac{\sigma - \eta}{1 - \sigma}}
 \label{eqn:quantitiesDemanded} 
\end{equation}

\begin{equation}
  p_i^{*} = \alpha \beta_i(\beta_i q_i)^{\frac{-1}{\sigma}} Q^{\frac{\eta -\sigma}{\eta \sigma} }   
 \label{eqn:willignestopay} 
\end{equation}

From this optimization procedure we derive an equation that describes the demanded quantity of one species in terms of the  consumption given by the other. From the first order condition we have:

\begin{equation}
  q_2 = \left(  \left(\frac{p_1}{\alpha \beta_1} (\beta_1 q_1)^{\frac{1}{\sigma}} \right)^{\frac{\eta(\sigma-1)}{\eta -\sigma}}  - (\beta_1 q_1)^{\frac{\sigma - 1}{\sigma}}  \right)^{\frac{\sigma}{\sigma-1}} (\beta_2)^{-1}
  \label{eqn:demandedQuantity} 
\end{equation}

The fishers maximization of profits and the utilities from the household,  allows us to find the optimal number of firms practicing each m\`{e}tier $k$ (Eq. \ref{eqn:numberOfFimrs1}, \ref{eqn:numberOfFimrs2}). The assumption of $\bar{i}=2$ and $\bar{k}=2$, holds in order to find a mathematical expression that can be generalized. With these components the model is described.

\begin{equation}
 n_1^* (q(\textbf{p})) = \frac {\nu_{22} \chi_2 q_1(\textbf{p}) - \nu_{12} \chi_1 q_2 (\textbf{p})}{ e_1^{*\epsilon} \chi_1 \chi_2 (\nu_{11} \nu_{22} - \nu_{12} \nu_{21})} 
\label{eqn:numberOfFimrs1}
\end{equation}
 
\begin{equation}
 n_2^* (q(\textbf{p})) = \frac {\nu_{11} \chi_1 q_2(\textbf{p}) - \nu_{12} \chi_2 q_1 (\textbf{p})}{ e_2^{*\epsilon} \chi_1 \chi_2 (\nu_{11} \nu_{22} - \nu_{12} \nu_{21})} 
\label{eqn:numberOfFimrs2}
\end{equation}

\section{APPENDIX}
\label{sec:appendix2}
\renewcommand{\theequation}{B.\arabic{equation}}
\setcounter{equation}{0}

\textbf{Model Calibration}

We calibrate stocks, harvests, and prices for the whole North Sea, using the data provided by the International Council for the Exploration of the Sea (ICES) regarding landings (harvests) and stocks (SSB)\citep{ICES2019, ICES2015}. Prices are calibrated using data from the EUMOFA (European Market Observatory for Fisheries and Aquaculture Products) database. The calibration involves the following steps:

\begin{enumerate}
\item \textit{Ecosystem component}: The function describing the stock growth is calibrated using data of SSB for plaice and sole from the years 1957 to 2019 \citep{ICES2019, ICES2015}. The data are transformed to a scale of the model through a scale parameter ($\kappa$) that represents the Maximum Sustainable Yield (MSY). The initial values of the parameters ‘a' and ‘b' of the Ricker-recruitment function are found by linearizing the function and fitting a linear model to the observed data (Equation.\ref{eqn:rickerfunction}), using the FSA (Simple Fisheries Stock Assessment Methods) library in R.	 Then, a nonlinear least squares model based on those values of ‘a' and ‘b' is estimated. The fit between the model estimates and the real growth data is shown in figure \ref{fig:ssb}.    

\item \textit{Household and Harvesting Components}: 
We calibrate household parameters using data prices for the years 2001-2020. We transform this prices to be relative to income to fit the scale of the model. We use the Gross Domestic Product (GDP) of the North Sea Countries as a proxy for the income used the model. Prices and GDP are adjusted for inflation to 2015 constant prices.
To calibrate harvesting parameters, we use the same data as in the step one, combined with harvest data reported in landings for the whole North Sea by \citet{ICES2021a, ICES2021b}\footnote{The \citet{ICES2021a, ICES2021b} reports include landings, discards and catches. For our purposes we set landings equivalent to harvest because these are the quantities that are traded on the market.}. Using this data and the parameters already found in step one we construct an objective function to minimize the error between the predicted values of harvests and prices, and the real data (Equation \ref{eqn:calibOptFun}). We use the existing implementation of the 
 \textit{nlminb} procedure in R to minimize these errors ($\zeta$) \citep{Nash2019}(Equation. \ref{eqn:harvestspermetier}, \ref{eqn:willignestopay}).

To find the initial values of our final calibration procedure we use results of previous trials. During our calibration procedure we implement different trials to minimize the objective function. Using different weighted values, including more or less parameters or changing the time lapse for the calibration. The result of these trials gives many possible values for each parameter. Use choose the maximum and minimum values of each parameter and construct a matrix of 519 possible combinations that we use as initial values for our final calibration.

Finally, we find the best fitting parameters for $\epsilon$, $\chi_1$, $\chi_2$,$\phi$, $\eta$, $\alpha$, $\sigma$, $\beta_1$, and $\beta_2$ that ensure an interior steady-state and reflects the real relationships between quantities, harvest and prices. To identify the parameters that comply with an interior steady-state we set the quota as the last value of our data for the year 2020.

\begin{equation}
 \min_{\hat{H_i},\hat{p_i} }  \zeta = \sum\limits_{t=1957}^{2020}\sum\limits_{i=1}^{2} m_i^h(\hat{H_{it}}- H_{it})^2 + \sum\limits_{t=2001}^{2020}\sum\limits_{i=1}^{2}m_i^p(\hat{p_{it}} - p_{it})^2 
 \label{eqn:calibOptFun}
\end{equation} 

 subject to: 

 \centerline {$\epsilon$, $\chi_1$, $\chi_2$,$\phi$, $\eta, \alpha,\beta_1$, and $\beta_2$ $>0.000001$ }
 \centerline {$\sigma$ $>1.000001$ }

where $\hat{H_{i}}$ is:

\centerline {$\hat{H_{it}} = \sum\limits_{k=1}^{2}  n_{kt}h_{ikt}(e_{kt}, x_{it}) =
 \sum\limits_{k=1}^{2} n_{kt}(\nu_{ik} e_{k}^ {\epsilon} x_{it}^{\chi_{i}}) $}

$\hat{p_{it}}$ is:

\centerline {$  \hat{p_{it}} = \alpha \beta_i(\beta_i q_{it})^{\frac{-1}{\sigma}} Q^{\frac{\eta -\sigma}{\eta \sigma} } $}

and $m_i^h$ and $m_i^p$ are weighted values for harvest and prices to normalize the calibration to the mean of the real values:

\centerline {$m_i^h = \frac{1}{\bar{H_i}}$ and $m_i^p = \frac{1}{\bar{p_i}}$  }

\end{enumerate}

\newpage

\section{APPENDIX}
\label{sec:appendix3}

\begin{figure}[h]
  \includegraphics[width=\linewidth]{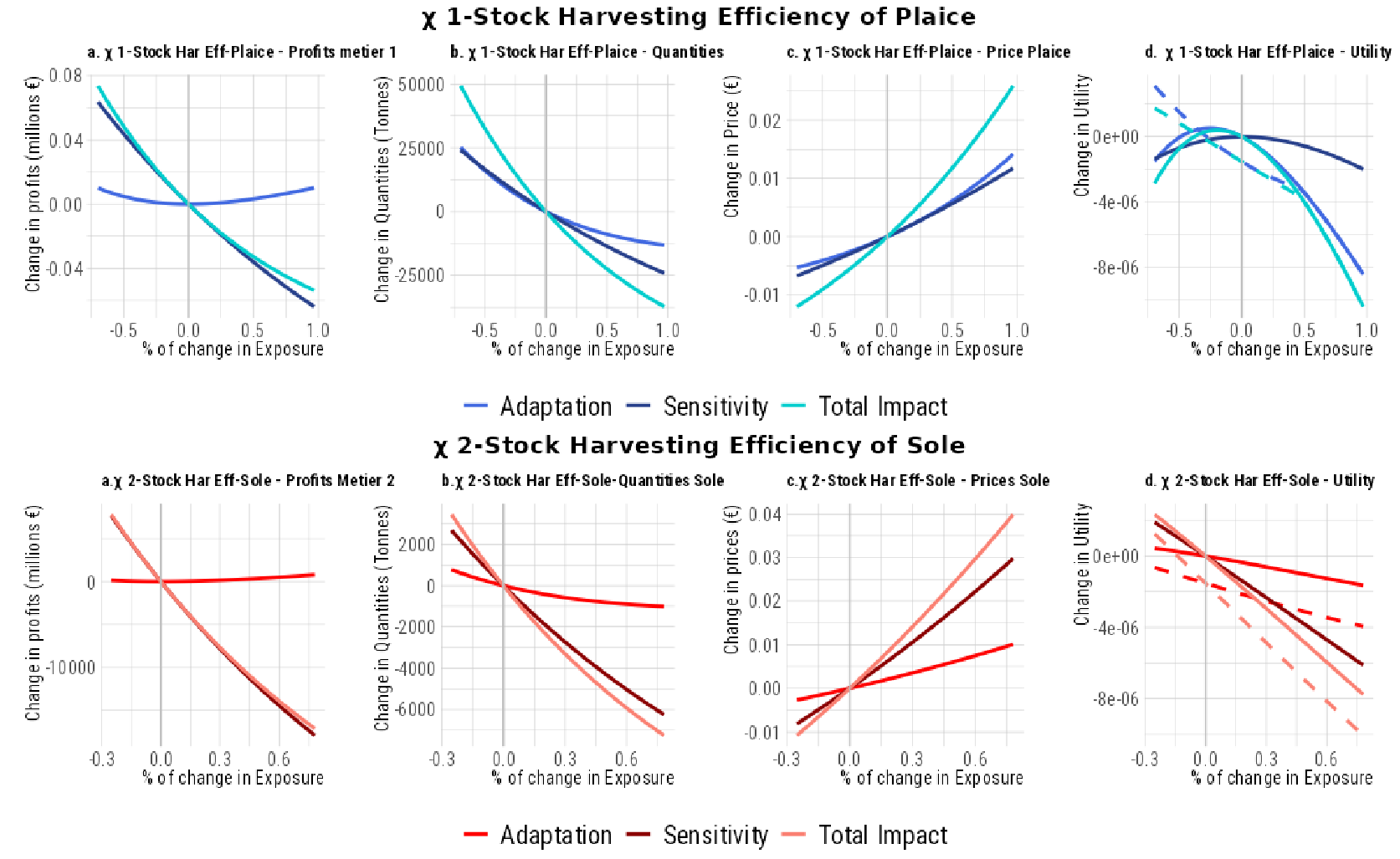}
  \caption {Changes in profits, quantities, prices and utility due to changes in stock harvesting efficiency.}
  \label{fig:chiAnalysis}
\end{figure}

\newpage

\begin{figure}[h]
  \includegraphics[width=\linewidth]{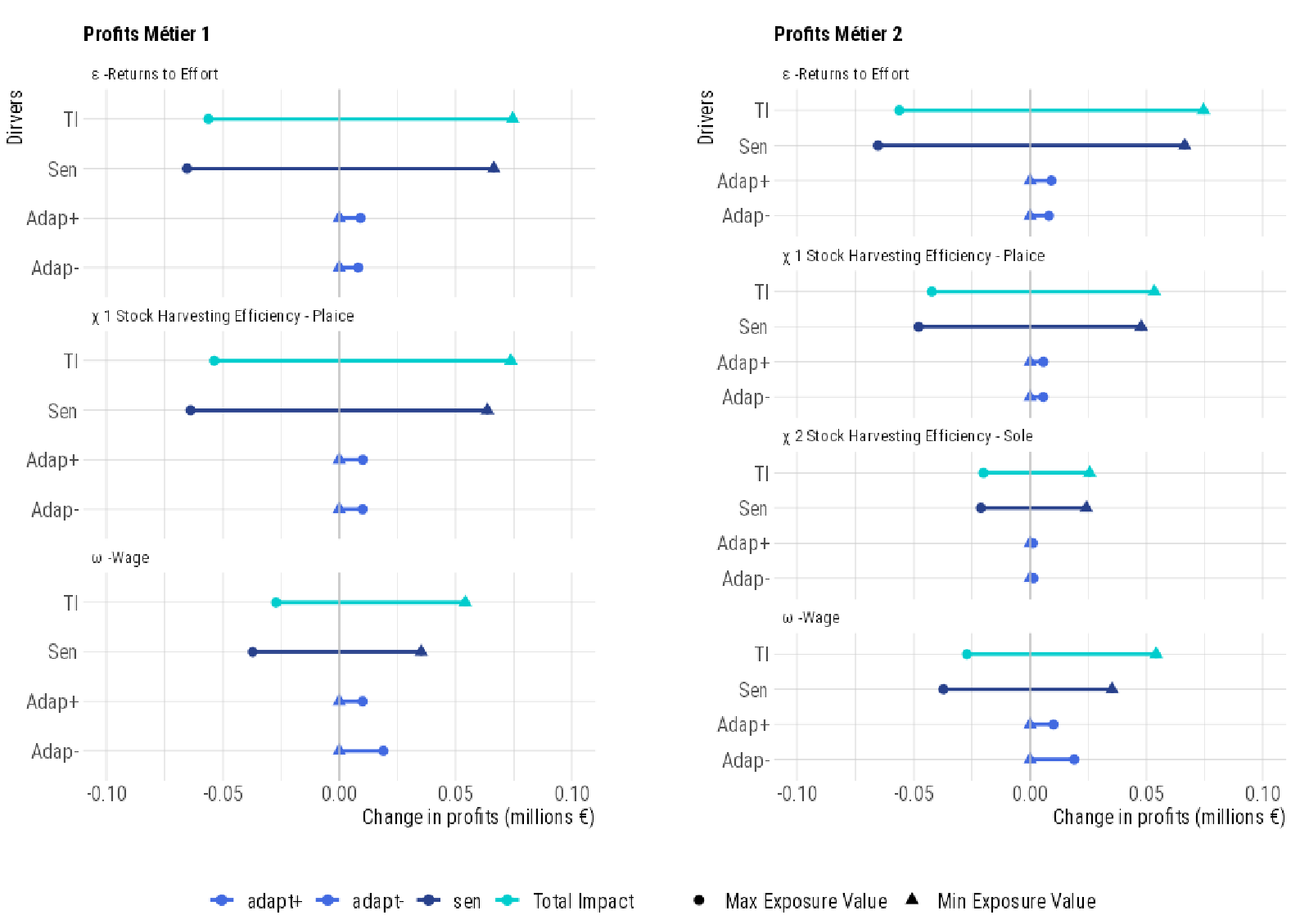}
  \caption {Absolute changes in profits due to changes in drivers.}
  \label{fig:overallProfits}
\end{figure}

\end{document}